\documentclass[a4paper,11pt]{article}
\usepackage{jheppub} 

\usepackage{amsmath,amssymb}
\usepackage{graphicx}
\usepackage{units}

\usepackage{braket}

\usepackage{comment}

\newcommand{\GeV}{\ \text{GeV}}

\newcommand{\be}{\begin{equation}}
\newcommand{\ee}{\end{equation}}
\newcommand{\eps}{\epsilon}
\newcommand{\ord}[1]{\mathcal{O}\left( #1 \right)}
\newcommand{\eref}[1]{(\ref{#1})}

\title{\boldmath Gravitational waves from cosmic strings
in Froggatt-Nielsen flavour models}

\author[a,b]{Simone Blasi,}
\affiliation[a]{Deutsches Elektronen-Synchrotron DESY, Notkestr. 85, 22607 Hamburg, Germany}
\affiliation[b]{Theoretische Natuurkunde and IIHE/ELEM, Vrije Universiteit Brussel, \& The International
Solvay Institutes, Pleinlaan 2, B-1050 Brussels, Belgium}

\author[c]{Lorenzo Calibbi,}
\affiliation[c]{School of Physics, Nankai University, Tianjin 300071, China}

\author[b]{Alberto Mariotti\,}

\author[b,d]{and Kevin Turbang\,}
\affiliation[d]{Universiteit Antwerpen, Prinsstraat 13, 2000 Antwerpen, Belgi\"e}

\emailAdd{simone.blasi@desy.de}
\emailAdd{calibbi@nankai.edu.cn}
\emailAdd{alberto.mariotti@vub.be}
\emailAdd{kevin.turbang@vub.be}

\abstract{
Gravitational waves (GWs) are a powerful probe of the earliest moments in the Universe, enabling us to test fundamental interactions at energy scales beyond the reach of laboratory experiments. In this work, we assess the GW capability to probe the origin of the flavour sector of the Standard Model (SM). 
Within the context of Froggatt-Nielsen models of fermion masses and mixing based on a gauged $U(1)$ flavour symmetry, we investigate the formation of cosmic strings and the resulting GW background (GWB), estimating the sensitivity to the model's parameter space of future GW experiments.
Comparing these results with the bounds from low-energy flavour observables, we find that these two types of experimental probes of the model are nicely complementary.
Flavour physics observables can probe low to intermediate symmetry-breaking scales $v_\phi$, while future GW experiments are sensitive to the opposite regime, for which the string tension is large enough to yield sizeable GW signals, and in the long run can set an upper limit on the scale as stringent as $v_\phi \lesssim 10^9$~GeV. 
In certain scenarios, the combination of flavour constraints and future GW bounds can bring about a complete closure of the available parameter space, which illustrates how GWB searches can play an important role in testing the origin of the SM flavour sector even if that occurs at ultra-high energies.
}

\preprint{
\begin{flushright}
DESY--24--147
\end{flushright}
}

\begin{document}

\maketitle
\flushbottom


\section{Introduction}

The Standard Model (SM) of particle physics provides a very successful description of the known fundamental particles and interactions, and has been tested with extreme accuracy at high-energy colliders and low-energy precision experiments.
However, there are still gaps in our understanding of 
high-energy physics. One fundamental feature of the SM that has been lacking a convincing explanation for decades is the existence of three families (or flavours) of fermions and the peculiar pattern of their masses and mixing. This is what is usually referred to as the ``flavour puzzle''.

In particular, the measured fermion masses span several orders of magnitudes and within the SM there is no profound explanation for these large hierarchies. Many ideas for theories beyond the Standard Model (BSM) have been proposed to solve this puzzle, see e.g.~\cite{Feruglio:2015jfa,Xing:2020ijf}. These models often involve the existence of new symmetries (global or gauged) beyond those of the SM. A paradigmatic (and among the earliest) example includes the Froggatt-Nielsen mechanism~\cite{Froggatt:1978nt,Leurer:1992wg,Leurer:1993gy}, where one or more $U(1)$ symmetries (or discrete subgroups thereof) are added to the SM such that their spontaneous breaking can lead to the natural generation of the fermion masses as higher-dimensional operators of an effective field theory (EFT) involving the SM fermions, the Higgs field, and the vacuum expectation value (vev) of one or more scalar fields that break the symmetry.

Even though models of this kind can leave imprints in some low-energy or TeV-scale observables -- e.g.~at colliders or flavour experiments~\cite{Tsumura:2009yf,Calibbi:2012at,Bauer:2016rxs,Ema:2016ops,Calibbi:2016hwq,Smolkovic:2019jow} -- they often rely on a new physics sector at very high energies, which is not directly testable in particle physics experiments. It is therefore desirable to find alternative ways to test the origin of the SM flavour sector.

Within this context, gravitational waves (GWs) provide a new way to test high-energy physics through the possible imprints that phenomena in the very early stage of the Universe, such as phase transitions or cosmic defects, could have left in terms of a gravitational-wave background (GWB) -- for an introduction to cosmic defects, see \cite{Vilenkin:2000jqa}. The detection of the GWB is indeed a major target of current and future GW experiments, see e.g.~\cite{Caldwell:2022qsj,LIGOScientific:2021nrg,NANOGrav:2023gor,NANOGrav:2023hvm,Xu:2023wog,EPTA:2023fyk,LISACosmologyWorkingGroup:2022jok} and references there in.

New symmetries spontaneously broken at very high temperatures, such as those often appearing in models addressing the flavour puzzle, lead to the formation of topological defects in the early Universe through the Kibble mechanism~\cite{Kibble:1976sj}. The networks of these defects (e.g.~cosmic strings or domain walls) are known to be powerful GWB sources~\cite{Vilenkin:2000jqa}. 

This suggests an interesting interplay between flavour observables and GWB signatures in testing models that address the flavour puzzle, potentially exploring complementary regions of parameter space. 

In order to investigate this interplay,
in this paper we consider a simple flavour model, based on an abelian gauged 
Froggatt-Nielsen (FN) symmetry
with appropriate charge assignment.
First, we review the basic setup (Section~\ref{sec:FN}), introduce the interactions of the FN gauge boson (Section~\ref{sec:FN-Zprime}), and discuss the existing constraints deriving from flavour observables (Section~\ref{sec:flavour}). 
Then, we study the resulting GW signal emerging from the string network generated at the spontaneous FN symmetry breaking phase transition (Section~\ref{sec:strings}).
In this discussion we take into account the fact that the strings can have a large width, modifying the GW spectrum with a UV cutoff that would affect its detectability in some regions of the parameter space. 
We then combine the two analyses to show how the flavour probes can be complementary to the GW signatures from cosmic strings in constraining the parameter space of flavour models based on new symmetries at high energies (Section~\ref{sec:results}).

To the best of our knowledge, a discussion on the GWB originating from cosmic strings in FN models has never been attempted. 
However, 
there are a few related works in the literature, which can be regarded as other instances of the fruitful interplay we would like to outline here.
In Ref.\,\cite{Baldes:2018nel} a first order phase transition at high temperatures associated to a scalar field giving mass to the FN messengers was considered as a potential source of GWs. The mechanism behind the dynamical generation of the SM Yukawas was shown to play a role in inducing a strongly first order electroweak phase transition in Ref.\,\cite{Baldes:2016gaf}.
In Ref.~\cite{Greljo:2019xan}, the existence of multiple peaks in the GWB spectrum is proposed as a possible signature of strong first order phase transitions (SFOPTs) associated with the breaking pattern of a model of flavour hierarchies involving three family-dependent copies of the Pati-Salam gauge group. Ref.~\cite{Gelmini:2020bqg} considered a model of neutrino masses and mixing based on the discrete symmetry $A_4$ acting on the lepton sector and discussed the GWB signal resulting from the annihilation of the domain walls produced after symmetry breaking -- see also~\cite{Jueid:2023cgp,Chauhan:2023faf,Wu:2022tpe,Fu:2024jhu} for further discussions on domain walls from discrete flavour symmetries.
Finally, Ref.~\cite{Ringe:2022rjx} explored the possibility of a SFOPT and the resulting GW signature within two global FN models similar to the (local) one we focus on in this work, finding that, for certain values of the parameters, the GWB can be strong enough to be detected in future experiments if the symmetry breaking occurs at an intermediate energy scale, $10^4-10^7$~GeV. As we will show, such a range is mostly excluded by flavour constraints within our setup. Furthermore, we do not investigate the possibility of a SFOPT here (as it typically requires the parameters of the model to satisfy quite non-trivial conditions) and focus on the GWB produced by the cosmic string network.

\section{Benchmark Froggatt-Nielsen Model}
\label{sec:FN}

The hierarchical structure of the Yukawa matrices can be accounted for by the FN mechanism~\cite{Froggatt:1978nt,Leurer:1992wg,Leurer:1993gy}. A new abelian symmetry $U(1)_F$ is introduced within this framework, under which SM fermions are charged such that the Yukawa interactions are forbidden at the renormalisable level (with the possible exception of the top quark Yukawa that, being $\ord{1}$, requires no suppression). The Yukawa couplings then arise as higher-order operators after the flavour symmetry is broken spontaneously by the vev of a complex scalar field $\phi$. This new field, also known as the ``flavon'', is not charged under any of the SM gauge symmetries and contains two degrees of freedom: a CP-even (real) scalar with mass $\ord{\langle\phi\rangle}$ and a CP-odd scalar, the Nambu-Goldstone boson of the broken $U(1)_F$, which becomes the longitudinal component of the associated gauge boson if the symmetry is local. 

As mentioned above, the mechanism requires that the SM fermions $f_i$ carry $U(1)_F$ charges $\mathcal{Q}_{f_i}$. Here, $f_i$ denotes the SM fermion fields with well-defined electroweak quantum numbers, with the generation index running over $i=1,2,3$ in the interaction basis. 

We consider the effective theory below a given UV cutoff scale $\Lambda$, which we take much higher than the electroweak scale. The flavon interacts with SM fields through higher-dimensional operators consistently with $U(1)_F$ invariance. Without loss of generality, one can set the flavon charge to be $\mathcal{Q}_\phi= 1$, obtaining the following interactions:
\begin{align}
-\mathcal{L} =
a_{ij}^u \left( \frac{\phi}{\Lambda} \right)^{n^u_{ij}} \overline{Q}_i u_j\, \tilde{H}\,  + 
a_{ij}^d \left( \frac{\phi}{\Lambda} \right)^{n^d_{ij}} \overline{Q}_i d_j\, H \,+  a_{ij}^\ell \left( \frac{\phi}{\Lambda_\ell} \right)^{n^\ell_{ij}} \overline{L}_i e_j\, H + {\rm h.c.} \ ,
\label{eq:lagr}
\end{align}
where the exponent of each term ensures the invariance of the  Lagrangian under $U(1)_F$,
\begin{align}
n^u_{ij} \equiv {\mathcal{Q}_{Q_i} -\mathcal{Q}_{u_j} + \mathcal{Q}_{H}}\,,\quad
n^d_{ij} \equiv {\mathcal{Q}_{Q_i} -\mathcal{Q}_{d_j} - \mathcal{Q}_{H}}\,,\quad 
n^\ell_{ij} \equiv {\mathcal{Q}_{L_i} -\mathcal{Q}_{e_j} - \mathcal{Q}_{H}}\,,
\end{align}
while $a^u$, $a^d$, and $a^\ell$ are anarchical matrices of $\ord{1}$ coefficients, which are related to the fundamental couplings of the underlying UV-complete theory. Plausible UV completions can be realised by heavy vector-like fermions or additional scalar doublets and singlets (the so-called ``FN messengers'') with mass of the order of the cutoff scale $\Lambda$ and $\mathcal{O}(1)$ couplings with the flavon and/or the SM fields~\cite{Calibbi:2012yj,Calibbi:2012at}.
Note that we considered the possibility of a different cutoff scale in the lepton sector, that is, $\Lambda_\ell \neq \Lambda$, an assumption that is consistent with the fact that the FN messenger fields in UV-complete models generally carry different quantum numbers in the quark and lepton sectors.

The SM Yukawa interactions arise dynamically upon spontaneous breaking of the $U(1)_F$ symmetry due to the flavon vev $\langle \phi \rangle$. Crucial quantities for this framework are the ratios between this vev and the UV cutoff scales:
\be
\eps \equiv \frac{\langle \phi \rangle}{\Lambda} < 1\,, \quad
\eps_\ell \equiv \frac{\langle \phi \rangle}{\Lambda_\ell} < 1\,.
\ee
In terms of these quantities, the SM quark and lepton Yukawa matrices then read 
\begin{align}
Y^u_{ij} =a^u_{ij} \,\eps^{\mathcal{Q}_{Q_i} -\mathcal{Q}_{u_j} + \mathcal{Q}_{H}}, \qquad
Y^d_{ij} = a^d_{ij}\, \eps^{\mathcal{Q}_{Q_i} -\mathcal{Q}_{d_j} - \mathcal{Q}_{H}}, \qquad 
Y^\ell_{ij} = a^\ell_{ij}\, \eps_\ell^{\mathcal{Q}_{L_i} -\mathcal{Q}_{e_j} - \mathcal{Q}_{H}}  \ . 
\end{align}
Under the assumption of anarchical $\ord{1}$ coefficients, the fermion hierarchies are solely due to powers of the small parameters $\eps$ and $\eps_\ell$, that is, the hierarchical structure is ultimately controlled by the FN charges we assign to the SM fields. Incidentally, note how the mechanism is not sensitive to the absolute scales $\langle \phi \rangle$ and $\Lambda_{(\ell)}$ but only on their ratio $\eps_{(\ell)}$. 

The Yukawa matrices can be diagonalised by bi-unitary transformations:
\begin{align}
\label{eq:mass-basis}
Y^f = V^{f\dag} \hat{Y}^f W^{f},~\qquad f= u,d,\ell \ ,
\end{align}
where $\hat{Y}^f$ are flavour-diagonal matrices, and $V^f$ and $W^f$ are unitary matrices corresponding to rotations of left-handed~(LH) and right-handed~(RH) fields, respectively. The size of their entries is approximately 
\begin{align}
\label{eq:rot}
& V^{u}_{ij} \sim \eps^{\left|\mathcal{Q}_{Q_i} -\mathcal{Q}_{Q_j} \right|},\quad 
 V^{d}_{ij} \sim \eps^{\left|\mathcal{Q}_{Q_i} -\mathcal{Q}_{Q_j} \right|},\quad 
V^{\ell}_{ij} \sim \eps^{\left|\mathcal{Q}_{L_i} -\mathcal{Q}_{L_j} \right|}, \nonumber \\
& W^{u}_{ij} \sim \eps^{\left|\mathcal{Q}_{u_i} -\mathcal{Q}_{u_j} \right|},\quad 
W^{d}_{ij} \sim \eps^{\left|\mathcal{Q}_{d_i} -\mathcal{Q}_{d_j} \right|}\,,\quad
W^{\ell}_{ij} \sim \eps^{\left|\mathcal{Q}_{e_i} -\mathcal{Q}_{e_j} \right|}\,.
\end{align}
Subsequently, the CKM mixing matrix can then be defined as
\begin{equation}
	V_\textsc{ckm}=V^{u}\,V^{d\,\dagger}\,.
\end{equation}
Hence, for $\mathcal{Q}_{Q_1} > \mathcal{Q}_{Q_2} > \mathcal{Q}_{Q_3}$ (an ordering justified by the observed mass hierarchy), the resulting entries of the CKM matrix are
\begin{align}
V_{us} \sim \eps^{\mathcal{Q}_{Q_1} -\mathcal{Q}_{Q_2}}\,,\quad
V_{ub} \sim \eps^{\mathcal{Q}_{Q_1} -\mathcal{Q}_{Q_3}}\,,\quad
V_{cb} \sim \eps^{\mathcal{Q}_{Q_2} -\mathcal{Q}_{Q_3}}\,,
\end{align}
from which the following general order-of-magnitude prediction (independent of the specific charge assignment) follows:
\begin{align}
V_{ub} \sim V_{us} \times V_{cb}\,,
\end{align}
which is in good agreement with experimental observations.

The setup described above is general. In the following, we will introduce benchmark models separately for quarks and leptons, leaving open the possibility that the FN symmetry only acts either on the quark sector or on the 
lepton sector, thus addressing the flavour hierarchies only partially, or assuming that different symmetries are at work in the two sectors.

Note that the FN charge of the Higgs field $H$ can always be taken to be vanishing since the FN charge of the full Yukawa operator determines the hierarchical suppression.\footnote{However, for a local FN symmetry, the Higgs field charge $\mathcal{Q}_{H}$ may make a difference: the Higgs kinetic term induces a mass mixing between the FN gauge boson and the SM $Z$ boson after electroweak symmetry breaking. Nevertheless, the mixing angle is suppressed by powers of ${v}/\langle \phi \rangle$ and thus negligible for a high-scale UV completion.}
In the remainder of this work, we will assume $\mathcal{Q}_{H} = 0$ for concreteness.

\subsection{Quark sector}

In the quark sector, a possible charge assignment is given by
\begin{align}
(\mathcal{Q}_{Q_1},\,\mathcal{Q}_{Q_2},\,\mathcal{Q}_{Q_3})&~=~(3,\,2,\,0),\nonumber\\
(\mathcal{Q}_{u_1},\,\mathcal{Q}_{u_2},\,\mathcal{Q}_{u_3})&~=~(-4,\,-2,\,0),\nonumber\\
(\mathcal{Q}_{d_1},\,\mathcal{Q}_{d_2},\,\mathcal{Q}_{d_3})&~=~(-4,\,-2,\,-2),
\end{align}
which leads to the following structure for the Yukawa matrices:
\begin{align}
Y^u \sim \left(\begin{array}{ccc}
\eps^7 & \eps^5 & \eps^3 \\
\eps^6 & \eps^4 & \eps^2 \\ 
\eps^4 & \eps^2 & 1
\end{array}\right),\quad
Y^d \sim \left(\begin{array}{ccc}
\eps^7 & \eps^5 & \eps^5 \\
\eps^6 & \eps^4 & \eps^4 \\ 
\eps^4 & \eps^2 & \eps^2
\end{array}\right).
\end{align}
Taking the expansion parameter of the order of the Cabibbo angle, 
$$\eps~\approx~0.2\,,$$ 
and given the freedom of choosing the $\ord{1}$ coefficients in $a^u$ and $a^d$,
the above matrices can easily fit the observed quark masses and CKM mixing. We stress that the discussion in the following sections depends only mildly on the specific values of the FN charges and could be readily adapted to other options.\footnote{See e.g.~Refs.~\cite{Fedele:2020fvh,Nishimura:2020nre,Cornella:2023zme} for recent fits of FN models to SM data and discussions of alternative/minimal charge assignments.} 

The order of magnitude of the rotations following from Eq.~\eref{eq:rot} is
\begin{align}
V^{u,d} \sim \left(\begin{array}{ccc}
1& \eps & \eps^3 \\
\eps &1 & \eps^2 \\ 
\eps^3 & \eps^2 & 1
\end{array}\right),\quad
W^u \sim \left(\begin{array}{ccc}
1 & \eps^2 & \eps^4 \\
\eps^2 & 1 & \eps^2 \\ 
\eps^4 & \eps^2 &1
\end{array}\right),\quad
W^d \sim \left(\begin{array}{ccc}
1 & \eps^2 & \eps^2 \\
\eps^2 & 1 & 1 \\ 
\eps^2 & 1 &1
\end{array}\right),
\end{align}
where we see that the rotations of the LH fields are of the order of the CKM angles both in the up and in the down sector.

\subsection{Lepton sector}

In the lepton sector, let us assume that neutrinos are Majorona particles, with their mass terms induced by the usual Weinberg operator~\cite{Weinberg:1979sa}. The resulting $U(1)_F$-invariant Lagrangian reads
\begin{align}
    -\mathcal{L}\supset \left[a^{\ell}_{ij}\, \left(\frac{\braket{\phi}}{\Lambda_\ell}\right)^{\mathcal{Q}_{L_i}-\mathcal{Q}_{e_j}}\overline{L}_{i}e_j H+h.c.\right]
    + \kappa^\nu_{ij}\,\left(\frac{\braket{\phi^*}}{\Lambda_\ell}\right)^{\mathcal{Q}_{L_i}+\mathcal{Q}_{L_j}}\frac{(\overline{L^c_i}\,\tilde H )(\tilde H^T L_j)}{\Lambda_N}\,.
    \label{eq:lepton-lag}
\end{align}
In the above Lagrangian, $\Lambda_N$ is the lepton number breaking scale, possibly related to the mass of heavy RH neutrinos. The SM Yukawa matrix and Majorana neutrino masses are dynamically generated in the following form:
\begin{equation}
    Y^\ell_{ij}=a^\ell_{ij}\,\epsilon_\ell^{\mathcal{Q}_{L_i}+\mathcal{Q}_{e_j}}\,, \quad\quad
    m^\nu_{ij}=\kappa^\nu_{ij}\,\frac{v^2}{\Lambda_N} \epsilon_\ell^{\mathcal{Q}_{L_i}+\mathcal{Q}_{L_j}}\,,\quad\quad \epsilon_\ell\equiv\frac{\braket{\phi}}{\Lambda_\ell}\,,
\end{equation} 
where the electroweak-breaking vev is defined as $\langle H \rangle = v/\sqrt2$.
As before, the elements of the matrices $a^\ell_{ij}$ and $\kappa^\nu_{ij}$ are assumed to be anarchical $\mathcal{O}(1)$ coefficients, related to the fundamental couplings of the underlying UV-complete theory. It follows that the hierarchy of lepton masses and mixing is solely due to powers of the $\epsilon_\ell$ parameter (possibly different from the expansion parameter $\epsilon$ of the quark sector), hence depending on the charges we assign to SM leptons. 

The above matrices can be diagonalised by means of unitary rotations of the fields:
\begin{align}
    Y^\ell=V^{\ell}\hat{Y}^\ell W^{\ell\dagger},\quad\quad
    m^\nu=V^{\nu}\hat{m}^\nu V^{\nu T}
\end{align}
where $\hat{Y}^\ell$ and $\hat{m}^\nu$ are flavour-diagonal matrices,
and the rotations to the mass basis have the following structure controlled by the FN charges:
\begin{equation}
\label{eq:rotations}
    V^{\ell,\nu}_{ij}\sim \epsilon_\ell^{\left|\mathcal{Q}_{L_i}-\mathcal{Q}_{L_j}\right|}, \;\;\;\;\;\; W^\ell_{ij}\sim \epsilon_\ell^{\left|\mathcal{Q}_{e_i}-\mathcal{Q}_{e_j}\right|}.
\end{equation} 
As usual, the PMNS matrix depends on the LH rotations as follows:
\begin{align}
\label{eq:pmns}
    U_\textsc{pmns} = V^{\nu} V^{\ell\,\dagger}.
\end{align}
In contrast to the quark sector, where the mixing is described by the strongly hierarchical CKM matrix, in the lepton sector the large mixing angles can be regarded as random $\mathcal{O}(1)$ numbers. In other words, given the current experimental precision, the observed mixing in the neutrino sector is compatible with an anarchical PMNS matrix~\cite{Hall:1999sn}. 
From Eqs.~(\ref{eq:rotations}) and (\ref{eq:pmns}), we see that this pattern can be simply achieved by taking equal charges for the doublets:
\begin{align}
\label{eq:anarchy}
    (\mathcal{Q}_{L_1},\,\mathcal{Q}_{L_2},\,\mathcal{Q}_{L_3})&~=~(\mathcal{Q}_{L},\,\mathcal{Q}_{L},\,\mathcal{Q}_{L}) & [\text{Pure Anarchy}]\,,
\end{align} 
as is also apparent from the above expression for $m^\nu_{ij}$.\footnote{Considering instead Dirac neutrinos would not change this conclusion, nor the the discussion below.}
In particular, it is interesting to note that we can choose to set $\mathcal{Q}_{L}=0$, such that the gauge boson $Z^\prime$ of the gauged $U(1)_F$ does not couple at tree-level to the LH leptons (including neutrinos). 
The hierarchy of the charged lepton masses can be reproduced by a suitable choice of the charges of the RH fields, the expansion parameter $\epsilon_\ell$, and the $\mathcal{O}(1)$ coefficients $a^\ell_{ij}$. For instance, in the purely anarchical scenario, the correct order of magnitude is achieved by choosing:
 \begin{align}
    (\mathcal{Q}_{e_1},\mathcal{Q}_{e_2}, \mathcal{Q}_{e_3})=(\mathcal{Q}_{L}-4,\mathcal{Q}_{L}-2,\mathcal{Q}_{L}-1), \;\;\;\;\;\;\;\;\;\;\;\; \epsilon_\ell \approx \epsilon^2 \approx 0.04.\label{chargeassign}
\end{align} 
Given the moderate value of the reactor angle, $\theta_{e3}\approx0.1$~\cite{Esteban:2020cvm}, 
good fits to the neutrino oscillation data can also be obtained in presence of mildly hierarchical charges at the price of rather large values of $\epsilon_\ell\approx 0.3 - 0.4$ \cite{Altarelli:2012ia,Bergstrom:2014owa}:
\begin{align}
\label{eq:mutau}
(\mathcal{Q}_{L_1},\,\mathcal{Q}_{L_2},\,\mathcal{Q}_{L_3})&~=~(\mathcal{Q}_{L}+1,\,\mathcal{Q}_{L},\,\mathcal{Q}_{L}) & 
[\mu\tau~\text{Anarchy}]\,,\\
(\mathcal{Q}_{L_1},\,\mathcal{Q}_{L_2},\,\mathcal{Q}_{L_3})&~=~(\mathcal{Q}_{L}+2,\,\mathcal{Q}_{L}+1,\,\mathcal{Q}_{L}) &
 [\text{Hierarchy}]\,. \label{eq:hie}
\end{align}
Again, the observed mass hierarchy of charged leptons can be achieved with a suitable choice of $\mathcal{Q}_{e_i}$. However, note that, in these last two cases, quite large values of the charges may be needed to obtain the required suppression. Furthermore, in these scenarios
the $Z^\prime$ unavoidably couples at least to some LH leptons and in particular to neutrinos, which makes its phenomenology more dependent on the details of the neutrino sector (including unknown properties thereof, such as the absolute neutrino mass). For these reasons, we will consider the anarchical charge assignment in Eqs.~(\ref{eq:anarchy}) and (\ref{chargeassign}) as our benchmark scenario for numerical considerations.

\subsection{Flavon interactions and decays}
\label{sec:flavon}

Let us write the phase and the radial excitation of the flavon field $\phi$ as
\begin{align}
\phi =\frac{v_\phi + \varphi}{\sqrt2} \,e^{i\,a/v_\phi}\,,
\label{eq:vev}
\end{align}
where we defined $v_\phi$ such that $\langle\phi\rangle \equiv v_\phi /\sqrt{2}$. 
For a local $U(1)_F$, 
the would-be Nambu-Goldstone boson $a$ provides the longitudinal component for the FN gauge boson $Z^\prime$.\footnote{In the global case, the Nambu-Goldstone boson of a spontaneously broken flavour symmetry is usually called ``familon''~\cite{Davidson:1981zd,Wilczek:1982rv,Reiss:1982sq,Davidson:1983fy,Chang:1987hz,Berezhiani:1990jj}. In the case of a FN $U(1)_F$ with colour anomaly, the field $a$ has also been dubbed ``flaxion'' or ``axiflavon'', because it automatically provides a solution to the strong CP problem, behaving like a QCD axion~\cite{Ema:2016ops,Calibbi:2016hwq}.} 
After $U(1)_F$ breaking, 
following from the quartic self-coupling $V(\phi)\supset \frac{\lambda_\phi}{4} |\phi|^4$,
the radial mode $\varphi$ (which we name ``flavon'', after the field $\phi$ itself) acquires a mass
\begin{align}
    m_\varphi^2 = \frac{1}{2}\lambda_\phi v_\phi^2\,.
\end{align}

Therefore, for perturbative values of the self-interaction, the flavon mass is at most of the order of the FN-breaking scale (which, as we will see, flavour processes constrain to be at least above $10^6-10^7$~GeV) and can only be much smaller than that scale if $\lambda_\phi\ll 1$.

Owing to the effective operators in~Eq.~\eref{eq:lagr}, the flavon couples to SM fermions as follows:
\begin{align}
    - \mathcal{L} = n^f_{ij}\, \frac{m^f_{ij}}{v_\phi} \,\overline{f}_i P_R f_j\,\varphi +\text{h.c.}\,,
    \label{eq:phicoupl}
\end{align}
where $f=u,d,\ell$ and the mass matrices $m^f = Y^f v/\sqrt2$. Note that, as a consequence of the dependence on the exponents $n^f_{ij}$, these interactions are by construction flavour-violating, i.e.,~the $\varphi$ couplings are not diagonal in the fermion mass basis. Indeed, rotating the fields according to Eq.~\eref{eq:mass-basis}, one obtains the following scalar and pseudoscalar currents:
\begin{align}
    - \mathcal{L} = C^f_{S\,\alpha\beta}\, \overline{f}_\alpha f_\beta \,\varphi+ C^f_{P\,\alpha\beta}\, \overline{f}_\alpha \gamma_5 f_\beta\,\varphi\,,
\label{eq:flavonL}
\end{align}
where the indices $\alpha,\beta$ denote mass eigenstates and 
\begin{align}
    C^{f}_{S\,\alpha\beta}& \equiv \left({V^{f}_{\alpha i}} \mathcal{Q}_{{f_L}_i} V^{f\,*}_{\beta i}-{W^{f}_{\alpha i}} \mathcal{Q}_{{f_R}_i} W^{f\,*}_{\beta i}\right) \frac{m_{f_\beta} + m_{f_\alpha}}{2v_\phi}\,,\\
    C^{f}_{P\,\alpha\beta}& \equiv \left({V^{f}_{\alpha i}} \mathcal{Q}_{{f_L}_i} V^{f\,*}_{\beta i} + {W^{f}_{\alpha i}} \mathcal{Q}_{{f_R}_i} W^{f\,*}_{\beta i}\right) \frac{m_{f_\beta} - m_{f_\alpha}}{2v_\phi}\,.
\end{align}
Note that we neglected flavon couplings with neutrinos induced by the term in Eq.~\eref{eq:lepton-lag} that gives rise to the Weinberg operator, as they are suppressed by the tiny neutrino masses.

The flavon decay widths into fermion pairs read:
\begin{align}
   \Gamma(\varphi \to f_\alpha\overline{f}_\beta) =&~\frac{N^f_c\,m_\varphi}{8\pi} \sqrt{\left(1-\frac{(m_{f_\alpha}+m_{f_\beta})^2}{m^2_\varphi} \right)\left(1-\frac{(m_{f_\alpha}-m_{f_\beta})^2}{m^2_\varphi} \right)}\, \times\nonumber\\ 
   &\left[ \left(1- \frac{(m_{f_\alpha}+m_{f_\beta})^2}{m_\varphi^2} \right) |C^{f}_{S\,\alpha\beta}|^2 +  \left(1- \frac{(m_{f_\alpha}-m_{f_\beta})^2}{m_\varphi^2} \right) |C^{f}_{P\,\alpha\beta}|^2 \right],
\end{align}
where $N^f_c$ is a colour factor (i.e., $N^{u,d}_c = 3$, $N^{\ell}_c =1$). 

The above decay rates are suppressed by a factor $\sim(m_f / v_\phi)^2$ and thus become negligible when $v_\phi \gg v$. As a consequence, for a heavy flavon, the three body decays $\varphi \to f_\alpha\overline{f}_\beta h$ involving a physical Higgs $h$ may become dominant. These modes stem from the dimension-5 operators
\begin{align}
    - \mathcal{L} = \frac{C^f_{S\,\alpha\beta}}{v}\, \overline{f}_\alpha f_\beta  \,\varphi \,h+ \frac{C^f_{P\,\alpha\beta}}{v}\, \overline{f}_\alpha \gamma_5 f_\beta\,\varphi \,h \,,
\end{align}
which are obtained by replacing the Higgs vev with its physical excitation in Eq.~\eref{eq:flavonL}. For the resulting decay rates (in the limit $m_\varphi \gg m_h$), we obtain:
\begin{align}
   \Gamma(\varphi \to f_\alpha\overline{f}_\beta h) \simeq~\frac{N^f_c\,m^3_\varphi}{4(4\pi)^3\,v^2} 
   \left( |C^{f}_{S\,\alpha\beta}|^2 + |C^{f}_{P\,\alpha\beta}|^2 \right) \,.
\end{align}
Note that effectively these partial widths scale as $m_\varphi^3 /v_\phi^2$, while the corresponding two-body decays scale as
$m_\varphi v^2 /v_\phi^2$. Therefore, the former indeed dominates if $m_\varphi \gg v$.

Finally, for local theories, 
if the process is kinematically open, the flavon can also efficiently decay into FN gauge bosons:
\begin{align}
   \Gamma(\varphi \to Z^\prime Z^\prime) =&~ \frac{1}{32\pi}\frac{m_\varphi^3}{v_\phi^2} \sqrt{1-4\frac{m_{Z^\prime}^2}{m_\varphi^2}}\left(1 -4\frac{m_{Z^\prime}^2} {m_\varphi^2}+ 12\frac{m_{Z^\prime}^2}{m_\varphi^2}\right)\,.
\end{align}   

Neglecting loop-induced couplings to SM gauge bosons, which also stem from the scale-suppressed interactions with fermions discussed above, the total decay width of a heavy flavon is then given by
\begin{align}
    \Gamma_{\varphi} = \Gamma(\varphi \to Z^\prime Z^\prime)+&\sum_{\alpha,\beta}
    \left[
    \Gamma(\varphi \to u_\alpha\overline{u}_\beta) +
    \Gamma(\varphi \to d_\alpha\overline{d}_\beta) +
    \Gamma(\varphi \to \ell_\alpha\overline{\ell}_\beta) \right] +\nonumber\\
    &\sum_{\alpha,\beta}
    \left[
    \Gamma(\varphi \to u_\alpha\overline{u}_\beta h) +
    \Gamma(\varphi \to d_\alpha\overline{d}_\beta h) +
    \Gamma(\varphi \to \ell_\alpha\overline{\ell}_\beta h)
    \right]\,.
    \label{totalwidth_phi}
\end{align}
In practice, if the decay mode $\varphi \to Z^\prime Z^\prime$ is not open, the flavon will mostly decay to the heaviest flavours that are kinematically accessible, in particular $\varphi \to t c (h)$ and $\varphi \to b \bar b (h)$.\footnote{Note that the flavon coupling to $t\bar t$  vanishes in the interaction basis, since $\mathcal{Q}_{Q_3} = \mathcal{Q}_{u_3} = 0$. Therefore, it is suppressed by a $t$-$c$ rotation in the mass eigenstate basis.} This is a consequence of the hierarchy of the interactions in Eq.~\eref{eq:phicoupl}, which follows the hierarchy of the SM Yukawas.

\section{The Froggatt-Nielsen $Z^\prime$}
\label{sec:FN-Zprime}

Henceforth, we focus on a model with a local flavour symmetry.
Note that the above benchmark charge assignments are anomalous. Nevertheless, gauge anomalies could be taken care of by the model's UV completion or within a dark sector of the theory. For explicit realisations of the former mechanism, see Refs.~\cite{Alonso:2018bcg,Smolkovic:2019jow,Bonnefoy:2019lsn}.
If the $U(1)_F$ is local, the theory will include a $Z^\prime$ gauge boson with mass
\begin{align}\label{eq:mzp}	
m_{Z^\prime}~=~\sqrt{2}\, g_F \langle \phi\rangle = g_F\, v_\phi\,,
\end{align}
where $g_F$ is the $U(1)_F$ gauge coupling.
Throughout the paper, we assume the kinetic mixing between the $U(1)_F$ and the hypercharge gauge bosons to be negligible.
Hence, the couplings of this flavoured $Z^\prime$ to the SM fields are only controlled by the FN charges and $g_F$. Given the pattern of charges needed to reproduce the observed Yukawas (shown in the previous section), the $Z^\prime$ will preferably couple to lighter generations.
Specifically, we have the following $Z^\prime$ interactions with quark and lepton fields before EW symmetry breaking: 
\begin{align}
 \label{eq:Zpcoupl}
\mathcal{L} ~=~ & g_F\,Z^\prime_{\mu} \,\left[\overline{u}_i \gamma^\mu (\mathcal{Q}_{Q_i} P_L + \mathcal{Q}_{u_i} P_R) u_i \,
+ \overline{d}_i \gamma^\mu (\mathcal{Q}_{Q_i} P_L + \mathcal{Q}_{d_i} P_R) d_i \,+ \right. \nonumber \\ 
&\left.\quad \quad\quad\overline{\ell}_i \gamma^\mu (\mathcal{Q}_{L_i} P_L + \mathcal{Q}_{e_i} P_R) \ell_i \,
+ \bar{\nu}_i\gamma^{\mu}\mathcal{Q}_{L_i}P_L \nu_i \right]\,.
\end{align}
After EW symmetry breaking, we rotate the fields from the interaction basis to the mass basis by means of the matrices in Eq.~\eref{eq:mass-basis} thus obtaining:
\begin{align}
\mathcal{L} ~=~  & g_F\,Z^\prime_\mu ~\left[  
\overline{u}_\alpha \gamma^\mu (C^{u}_{L\,\alpha\beta} \,P_L +C^{u}_{R\,\alpha\beta} \,P_R) u_\beta\,  + \overline{d}_\alpha \gamma^\mu (C^{d}_{L\,\alpha\beta} \,P_L +C^{d}_{R\,\alpha\beta} \,P_R) d_\beta  \, + \right. \nonumber \\
& \left. \quad \quad\quad~ \overline{\ell}_\alpha \gamma^\mu (C^{\ell}_{L\,\alpha\beta}\,  P_L +C^{\ell}_{R\,\alpha\beta} \,P_R) \ell_\beta\, + \bar{\nu}_\alpha\gamma^{\mu} \, C^{\nu}_{L\,\alpha\beta} \,P_L \nu_\beta \right]\,,
 \label{eq:massbasis}
\end{align}
where the (hermitian) coupling matrices read:
\begin{align}
C^{f}_{L\,\alpha\beta} \equiv {V^{f}_{\alpha i}} \mathcal{Q}_{{f_L}_i} V^{f\,*}_{\beta i}\,,\quad\quad 
C^{f}_{R\,\alpha\beta} \equiv {W^{f}_{\alpha i}} \mathcal{Q}_{{f_R}_i} W^{f\,*}_{\beta i}\,.
\end{align}
In terms of couplings to vector and axial currents, we can then write:
\begin{align}
\mathcal{L} ~=~  & g_F\,Z^\prime_\mu ~\left[  
\overline{f}_\alpha \gamma^\mu (C^{f}_{V\,\alpha\beta} +C^{f}_{A\,\alpha\beta} \,\gamma_5) f_\beta \right] \,,
\quad
C^{f}_{V,A} = \frac{C^{f}_{R}\pm C^{f}_{L}}{2}\,.
\label{eq:CVA}
\end{align}

As we can see from Eq.~(\ref{eq:massbasis}), flavour-violating $Z^\prime$ couplings are typically induced because different flavours carry different $U(1)_{F}$ charges. In fact, because of the unitarity of the rotation matrices, 
flavour-violating couplings are proportional to the difference of the charges of the fermions involved.
As a consequence, our $Z^\prime$ generally mediates flavour-changing-neutral-current (FCNC) and lepton-flavour-violating (LFV) processes such as \mbox{$K-\bar K$} oscillations, $\mu\to eee$, etc. 
If $Z^\prime$ is light enough (which may occur for $g_F \ll 1$), 
one should also consider constraints from decays of mesons and leptons \emph{into} $Z^\prime$, such as 
$K\to \pi Z^\prime$ and $\mu\to e Z^\prime$. The resulting flavour bounds are discussed in the next section.

In terms of the couplings defined in Eq.~\eref{eq:CVA}, the kinematically allowed decay widths of the $Z^\prime$ decaying into fermions read~\cite{Kang:2004bz}:  
\begin{align}
      \Gamma(Z^\prime \to f_\alpha\overline{f}_\beta)&  =~\frac{N^f_c\,g_F^2\,m_{Z^\prime}}{12\pi} \sqrt{\left(1-\frac{(m_{f_\alpha}+m_{f_\beta})^2}{m^2_{Z^\prime}} \right)\left(1-\frac{(m_{f_\alpha}-m_{f_\beta})^2}{m^2_{Z^\prime}} \right)}  ~\times  \\
   & \Bigg[ \Bigg(1- \frac{m_{f_\alpha}^2+m_{f_\beta}^2}{2m_{Z^\prime}^2} \Bigg) \left(|C^{f}_{V\,\alpha\beta}|^2+|C^{f}_{A\,\alpha\beta}|^2\right) +  3\,\frac{m_{f_\alpha}m_{f_\beta}}{m_{Z^\prime}^2} \left(|C^{f}_{V\,\alpha\beta}|^2-|C^{f}_{A\,\alpha\beta}|^2\right) \Bigg], \nonumber
\end{align}   
where the colour factor is $N^{u,d}_c = 3$, $N^{\ell,\nu}_c =1$. 
In particular, the flavour-conserving decay widths take the form:
\begin{align}
    \Gamma(Z^\prime \to f_\alpha\overline{f}_\alpha)
     = \frac{N^f_c\,g_F^2\,m_{Z^\prime}}{12\pi}
    \sqrt{1-4\frac{m_{f_\alpha}^2}{m_{Z^\prime}^2}}
    \bigg[ 
    \bigg(1+2\frac{m_{f_\alpha}^2}{m_{Z^\prime}^2}  \bigg)|C^{f}_{V\,\alpha\alpha}|^2+\bigg(1-4\frac{m_{f_\alpha}^2}{m_{Z^\prime}^2}  \Bigg)|C^{f}_{A\,\alpha\alpha}|^2
    \bigg].
\end{align}
For a heavy $Z^\prime$, the total decay width is then just given by\,\footnote{A coupling of the $Z^\prime$ with photons is induced via fermion loops. However, according to the Landau-Yang theorem~\cite{Landau:1948kw,Yang:1950rg}, a vector boson cannot decay into two photons, which leaves $Z^\prime \to \gamma\gamma\gamma$ as the leading $Z^\prime$ decay into photons. This mode is highly suppressed and only relevant if $Z^\prime$ is lighter than any fermion pair it couples to~\cite{Redondo:2008ec} -- in our case, $m_{Z^\prime}< 2 m_{\nu_1}$, or $m_{Z^\prime}< 2m_e$ for models featuring no interaction with neutrinos, that is, $\mathcal{Q}_{L_i}=0$.}
\begin{align}
\label{eq:GammaZp}
    \Gamma_{Z^\prime} = \sum_{\alpha,\beta}
    \left[
    \Gamma(Z^\prime \to u_\alpha\overline{u}_\beta) +
    \Gamma(Z^\prime \to d_\alpha\overline{d}_\beta) +
    \Gamma(Z^\prime \to \ell_\alpha\overline{\ell}_\beta) +
    \Gamma(Z^\prime \to \nu_\alpha\overline{\nu}_\beta) 
    \right]\,.
\end{align}
For simplicity, when considering a light $Z^\prime$, we still estimate its lifetime  based on the above perturbative processes, neglecting hadronization and just eliminating the contributions below threshold. For what concerns light quarks, no contribution from decays into up and down (strange) quarks is included for $m_{Z^\prime}$ below the pion (kaon) kinematic threshold.

\section{Flavour constraints}
\label{sec:flavour}

In this section, we focus on the most relevant constraints from low-energy processes on FN models, which are due to the flavour-violating interactions of the FN gauge boson. In principle, the flavon $\varphi$ can also mediate flavour-changing processes. However, its contributions are suppressed by powers of $m_f/v_\phi$, as shown in Eq.~\eqref{eq:phicoupl}, hence resulting subdominant compared to the $Z^\prime$ contributions, see e.g.~\cite{Leurer:1992wg,Calibbi:2015sfa}. If $\varphi$ is so light that mesons or leptons can decay into it, such processes have the same parametric dependence as those involving a light $Z^\prime$ discussed below (see for instance~\cite{MartinCamalich:2020dfe}), such that the bound on $v_\phi$ set by the latter processes would only increase by an $\ord{1}$ factor. 

Other possible constraints on our parameter space, e.g., from the supernova explosion SN1987A and beam dump experiments, are less stringent than the flavour constraints, as shown in Ref.~\cite{Smolkovic:2019jow}. Therefore, we will not discuss them here.

\subsection{$Z^\prime$-mediated flavour violation}
As shown above, in general, flavour-violating $Z^\prime$ couplings are induced because FN models require that different flavours carry different FN charges by construction.
For instance, integrating out the $Z^\prime$, the Wilson coefficients of the dimension-6 $\Delta S=2$ \mbox{$(V-A)(V-A)$} and \mbox{$(V-A)(V+A)$}  operators in the down sector are (at leading order in our expansion parameter $\epsilon$):
\begin{align}
(\overline{s}_L \gamma^\mu d_L)( \overline{s}_L \gamma_\mu d_L):&\quad \frac{g_F^2}{m^2_{Z^\prime}}  \left[ (\mathcal{Q}_{Q_1} -\mathcal{Q}_{Q_2}) V^{d}_{21} \right]^2 \,, \\
(\overline{s}_L \gamma^\mu d_L)( \overline{s}_R \gamma_\mu d_R):&\quad \frac{g_F^2}{m^2_{Z^\prime}}  (\mathcal{Q}_{Q_1} -\mathcal{Q}_{Q_2})  
V^{d}_{21} (\mathcal{Q}_{d_1} -\mathcal{Q}_{d_2}) W^{d}_{21}\,.
\end{align}
Note that these Wilson coefficients are suppressed by a factor ${g_F^2}/{m^2_{Z^\prime}}= 1/v_\phi^2$, effectively depending on a single parameter: the FN symmetry breaking scale (besides a mild dependence on the model-dependent charge assignment). 
The above coefficients are strongly constrained by $K-\bar{K}$ mixing (that is, the kaon mass splitting $\Delta M_K$ and the CPV observable $\epsilon_K$), which provides the most important flavour bound on the mass/coupling of the $Z^\prime$.
Of course, this EFT approach is only consistent if $m_{Z^\prime}\gg m_K$.\footnote{The contributions to meson mixing of a light $Z^\prime$ can be found in Ref.~\cite{Smolkovic:2019jow}. However, they give rise to subdominant constraints compared to those coming from the meson decays \emph{into} $Z^\prime$ that are kinematically open in such a regime, hence we can safely neglect them.}

Comparing the coefficients of the $\Delta S=2$ operators with recent limits reported in the literature~\cite{Silvestrini:2018dos}, we find that the most stringent constraints
on the $Z^\prime$ mass arise from its contribution to $(\overline{s}_L \gamma^\mu d_L)( \overline{s}_R \gamma_\mu d_R)$. Neglecting the effect of the $\ord{1}$ coefficients encoded in the rotation matrices $V^d$ and $W^d$, we obtain the following order of magnitude estimates:
\begin{align}
\label{eq:DeltaMK}
\Delta M_K:~m_{Z^\prime} \gtrsim&~ 0.65\,\left[\frac{g_F}{10^{-6} } \right] \GeV  \quad
\Longleftrightarrow \quad v_\phi \gtrsim 6.5\times 10^5\GeV\,,  \\
\epsilon_K:~m_{Z^\prime} \gtrsim&~ 13\,\left[\frac{g_F}{10^{-6} } \right] \left[\frac{\alpha}{\ord{1}}\right] \GeV\quad  \Longleftrightarrow \quad v_\phi /\alpha \gtrsim 1.3\times 10^7 \GeV\,, 
\end{align}
where $\alpha \equiv  \sqrt{\arg \left(  V^{d}_{12}  W^{d}_{12} \right) }$ is a CPV phase.
Analogous $\Delta F = 2$ processes, i.e., $B_{(s)}-\bar{B}_{(s)}$ and $D-\bar D$ oscillations, set subdominant constraints on $v_\phi$ owing to the comparatively weaker limits on the coefficients of the corresponding operators~\cite{Silvestrini:2018dos}.

Similarly, in the lepton sector, LFV decays such as $\mu \to eee$ and $\mu\to e$ conversion in nuclei are induced at tree level. These processes are already tightly constrained by past experiments and an increased sensitivity of several orders of magnitude is expected at upcoming searches~\cite{Calibbi:2017uvl}.
If $m_{Z^\prime}\gg m_\mu$ the effect is described by
 effective operators of the following kind:
\begin{align}
(\overline{\mu}_R \gamma^\mu e_R)( \overline{e}_R \gamma_\mu e_R):&\quad \frac{g_F^2}{m^2_{Z^\prime}}  \mathcal{Q}_{e_1} 
(\mathcal{Q}_{e_1} -\mathcal{Q}_{e_2}) W^{e}_{21}\,,\\
(\overline{\mu}_R \gamma^\mu e_R)( \overline{q} \gamma_\mu q):&\quad \frac{g_F^2}{m^2_{Z^\prime}}  \mathcal{Q}_{q} 
(\mathcal{Q}_{e_1} -\mathcal{Q}_{e_2}) W^{e}_{21}\,,
\end{align}
where $q = u_{L,R},\,d_{L,R}$ and we do not consider LFV LH current operators as they vanish in the purely anarchical case given by Eq.~(\ref{eq:anarchy}).
For the charge assignment shown in Eq.~\eref{chargeassign} (with $\mathcal{Q}_L = 0$) the current limits on LFV operators (see e.g.~\cite{Calibbi:2017uvl}) from \mbox{BR$(\mu^+\to e^+e^-e^+) <10^{-12}$}~\cite{SINDRUM:1987nra} and CR$(\mu^-\,\rm{Au}\to e^-\,\rm{Au}) < 7\times 10^{-13}$~\cite{SINDRUMII:2006dvw} imply the following constraints on $v_\phi$:
\begin{align}
\mu\to eee:~m_{Z^\prime} \gtrsim& ~24\,\left[\frac{g_F}{10^{-3} }\right]\GeV\quad
\Longleftrightarrow \quad v_\phi  \gtrsim  2.4\times 10^4\,\GeV\,,\\
\mu\to e~\text{in N}:~m_{Z^\prime} \gtrsim& ~89\,\left[\frac{g_F}{10^{-3} }\right]\GeV\quad
\Longleftrightarrow \quad v_\phi  \gtrsim  8.9\times 10^4\,\GeV\,.
\end{align}
As we can see, if the same $U(1)_F$ symmetry is responsible for the hierarchies in both quark and lepton sectors, the most stringent limit on the breaking scale $v_\phi$ is set by $K-\bar K$ observables. In the following sections, we will adopt Eq.~\eref{eq:DeltaMK} as the limit on $v_\phi$ in the regime $m_{Z^\prime} > m_B$, conservatively assuming that a small CPV phase in the 1-2 quark rotations somewhat suppresses the $Z^\prime$ contribution to $\epsilon_K$.

\subsection{Decays into light $Z^\prime$}

If $Z^\prime$ is light -- that is, $g_F\ll 1$,  $m_{Z^\prime} \ll \langle \phi\rangle$ -- flavour-violating meson or lepton decays into an on-shell $Z^\prime$ itself (such as $B\to K Z^\prime$, \mbox{$K\to \pi Z^\prime$}, $\mu\to eZ^\prime$ etc.)~can be kinematically open and set the strongest constraints on the FN breaking scale. 

In the case of the meson decays $K^+ \to \pi^+ Z^\prime$ and $B^+ \to K^+ Z^\prime$, we have~\cite{Smolkovic:2019jow,Altmannshofer:2023hkn}:
\begin{align}
\text{BR}(K^+\to \pi^+ Z^\prime) = &~ \frac{g_F^2}{16\pi \,\Gamma_K} \frac{m^3_K}{m_{Z^\prime}^2} \,\left[\lambda\left(1,\frac{m_\pi^2}{m_K^2},\frac{m_{Z^\prime}^2}{m_K^2}\right)\right]^{\frac32} [f_+(m_{Z^\prime}^2)]^2 |C^{d}_{V\,s d}|^2\,,\\ 
\text{BR}(B^+\to K^+ Z^\prime) = &~ \frac{g_F^2}{16\pi \,\Gamma_B} \frac{m^3_B}{m_{Z^\prime}^2}\, \left[\lambda\left(1,\frac{m_K^2}{m_B^2},\frac{m_{Z^\prime}^2}{m_B^2}\right)\right]^{\frac32} [f_+(m_{Z^\prime}^2)]^2 |C^{d}_{V\,b s}|^2\,,
\end{align}
where the meson widths are $\Gamma_K \simeq 5.3\times 10^{-17}$~GeV and $\Gamma_B \simeq 4.0\times 10^{-13}$~GeV~\cite{PDG}, 
the kinematic factor is defined as $\lambda(x,y,z)\equiv x^2+y^2+z^2 - 2(xy+yz+xz)$, 
and the form factors $f_+$ are to be evaluated at $q^2= m_{Z^\prime}^2$.\footnote{For $m_{Z^\prime}\ll m_{K,B}$, this results in $f_+(0)\simeq 0.97$~\cite{Carrasco:2016kpy} for $s\to d$ transitions, and $f_+(0)\simeq 0.335$~\cite{Gubernari:2018wyi} for $b\to s$.} 
For our benchmark charge assignment the couplings in Eq.~\eref{eq:CVA} result $|C^{d}_{V\,s d}| \approx (\mathcal{Q}_{Q_1} -\mathcal{Q}_{Q_2}) V^{d}_{21}/2 \approx \epsilon/2$, $|C^{d}_{V\,b s}| \approx (\mathcal{Q}_{Q_2} -\mathcal{Q}_{Q_3}) V^{d}_{32}/2 \approx \epsilon^2$.

Similarly, the branching ratio of the leptonic decays $\ell_\alpha\to\ell_\beta Z^\prime$ reads~\cite{Heeck:2016xkh,Smolkovic:2019jow,Ibarra:2021xyk} (under the approximation $m_{\ell_\beta} \ll m_{\ell_\alpha}$):
\begin{align}
\text{BR}(\ell_\alpha\to \ell_\beta Z^\prime) =&~ \frac{g_F^2}{16\pi \,\Gamma_{\ell_\alpha}} \frac{m_{\ell_\alpha}^3}{m_{Z^\prime}^2} \left( |C^{\ell}_{V\,\alpha\beta}|^2 +
|C^{\ell}_{A\,\alpha\beta}|^2 \right)\left(1+2\frac{m_{Z^\prime}^2}{m_{\ell_\alpha}^2} \right)\left(1-\frac{m_{Z^\prime}^2}{m_{\ell_\alpha}^2} \right)^2 \,,
\end{align}
where for the total lepton widths we use $\Gamma_\mu \simeq m_\mu^5 G_F^2/(192\pi^3)$, $\Gamma_\tau \simeq 2.3\times 10^{-12}~\text{GeV}$~\cite{PDG}. The axial/vector couplings are defined in Eq.~\eref{eq:CVA}. Using again the charge assignment in Eq.~\eref{chargeassign} with $\mathcal{Q}_L = 0$, i.e., a $Z^\prime$ with purely RH couplings, we get
$C^{\ell}_{V\,\mu e} = C^{\ell}_{A\,\mu e}  \approx 
(\mathcal{Q}_{e_1} -\mathcal{Q}_{e_2}) W^{e}_{21}/2 \approx \epsilon_\ell^2 \approx \epsilon^4$.

The relevant experimental searches depend on the $Z^\prime$ lifetime. For a given value of $m_{Z^\prime}$, depending on the coupling $g_F$, $Z^\prime$ may indeed either decay into lepton pairs inside the detector or be long-lived enough to escape it, thus giving rise to a missing energy signature. 

In the latter case, one can compare the above rates with the limits from searches for an invisible boson $X$ in kaon decays at NA62~\cite{NA62:2021zjw} and in $B$ decays at $B$-factory experiments (as recast in Ref.~\cite{MartinCamalich:2020dfe} from searches for $B\to K\nu\bar\nu$):  for $m_X \ll m_{K/B}$, these limits are respectively $\text{BR}(K^+\to \pi^+ X) < 5\times 10^{-11}$, $\text{BR}(B^+\to K^+ X) < 7.1\times 10^{-6}$.

In the lepton sector, limits on $\tau\to \ell X$ have been recently published by Belle~II~\cite{Belle-II:2022heu} which, in the light $X$ limit, read
$\text{BR}(\tau\to e X) < 8.5\times 10^{-4}$, $\text{BR}(\tau\to \mu X) < 6\times 10^{-4}$. 
For what concerns the muon decay,
since the relevant experimental searches used polarised beams, the limit on $\text{BR}(\mu\to e X)$ depends on the angular distribution of the signal, ranging from $5.8\times 10^{-5}$ for a (practically) massless boson coupling mainly to LH leptons (thus to a $V-A$ current) to $2.5\times 10^{-6}$ if the couplings to RH leptons (hence to a $V+A$ current) dominate~\cite{Jodidio:1986mz,Calibbi:2020jvd}, like in our benchmark scenario. For what concerns heavier bosons, the dependence on $m_X$ is mild in the $V+A$ case and the average upper bound is $6\times 10^{-6}$~\cite{TWIST:2014ymv}, which is the limit that we employ in the following.

For definiteness, we apply the above limits to the portion of the parameter space where the $Z^\prime$ decay length $c\tau_{Z^\prime}  > 1$~m, given the typical size of the experimental apparatuses, with the exception of NA62 for which we take $c\tau_{Z^\prime}  > 100$~m, where $\tau_{Z^\prime} = 1/\Gamma_{Z^\prime}$ and the total $Z^\prime$ width is calculated considering the kinematically open decay modes in Eq.~\eref{eq:GammaZp}.
For larger values of $g_F$, the lifetime becomes shorter, such that the $Z^\prime$ would decay into lepton pairs inside the detector. In such a case, we can then impose constraints from $K \to \pi \ell\ell$,
$B \to \pi \ell\ell$, $\mu \to eee$, $\tau\to \mu \ell\ell$, where $\ell\ell = e^+e^- / \mu^+\mu^-$.\footnote{Our benchmark charge assignment yields $\text{BR}(Z^\prime\to \mu e) \lesssim 10^{-7}$ for the $Z^\prime$ LFV decay, which is too suppressed to give rise to relevant constraints from e.g.~searches for $K\to \pi e\mu$ or $B\to K e\mu$.}
Lepton decays into $Z^\prime$ are thus constrained by the above mentioned limit on $\mu^+ \to e^+ e^+ e^-$ and by searches for LFV 3-body $\tau$ decays at $B$ factories, which set limits in the $\mathcal{O}(10^{-8})$ range~\cite{PDG}.
For what concerns the meson decays, in view of the difficulties that affect the SM predictions (owing to long-distance effects) and/or some mild discrepancies with the data (in the case of the $B$ decays), we adopt for concreteness the conservative limits $\text{BR}(K \to \pi \ell\ell)< 10^{-7},\,\text{BR}(B \to K \ell\ell) < 10^{-7}$, which correspond to the order of magnitude of the measured branching fractions~\cite{PDG}.

\begin{figure}[t!]
\centering
\includegraphics[width=0.49\textwidth]{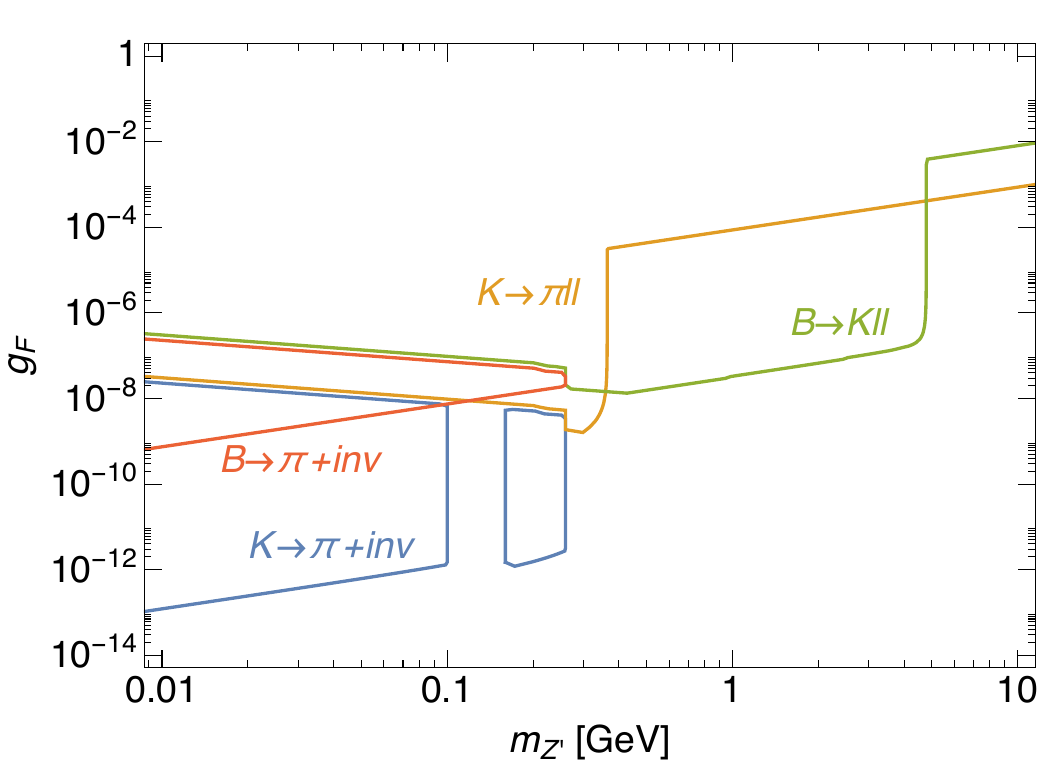}
\hfill
\includegraphics[width=0.49\textwidth]{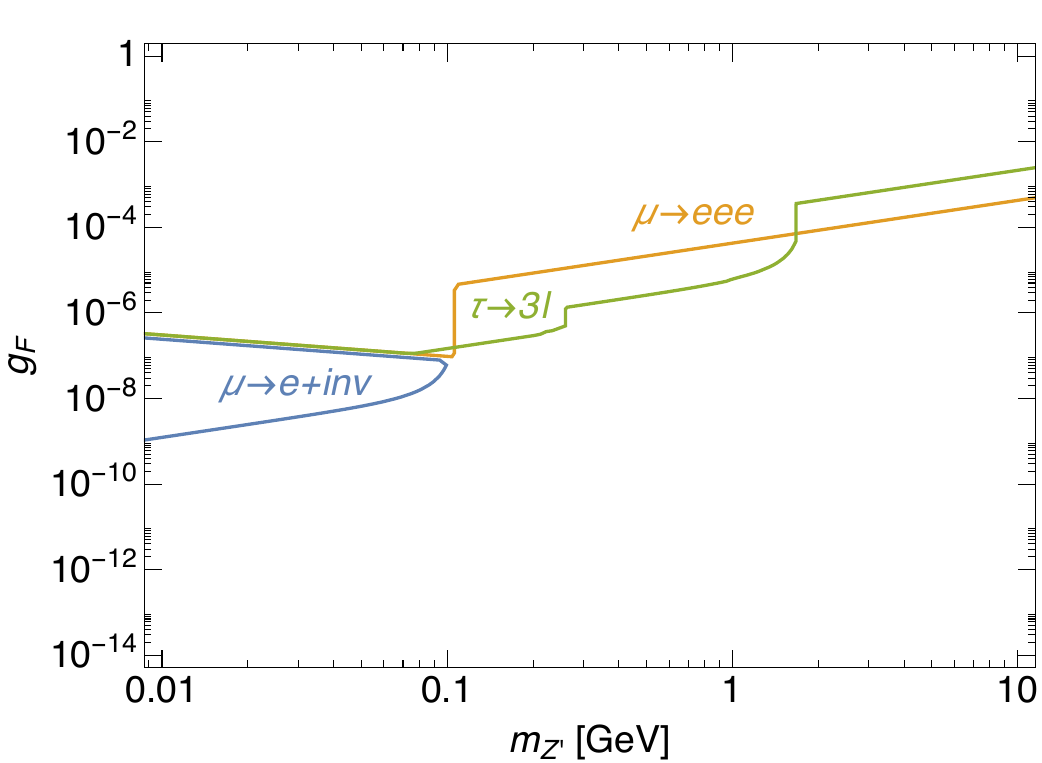}
\caption{Regions of the $m_{Z^\prime}-g_F$ plane excluded by searches for various rare meson (left) and lepton (right) decays into a light $Z^\prime$ (including the off-shell $Z^\prime$ case). See the main text for details.
}
\label{fig:flavour-bounds}
\end{figure}

The impact of these constraints on the parameter space featuring a light $Z^\prime$ is summarised in Figure~\ref{fig:flavour-bounds} separately for meson (left plot) and lepton (right plot) processes. Clearly the former supersede the latter if the same $U(1)_{F}$ is responsible for both quark and lepton hierarchies. We also note that searches for $B$ and $\tau$ decays into an invisible boson have little or no impact because the $Z^\prime$ lifetime is too short in the relevant ranges of $m_{Z^\prime}$, where the visible counterparts of these decays set the strongest bounds. 
As we can see, invisible and visible decays are complementary in constraining wide ranges of $g_F$ for a given $m_{Z^\prime}$, such that their combination yields the following approximate lower bounds on the $U(1)_F$ breaking scale in the relevant $m_{Z^\prime}$ ranges: 
\begin{align}
 K^+\to \pi^+ Z^\prime:&~~
v_\phi \gtrsim ~ 8.3\times 10^{10}~\text{GeV}\,,\quad
& B^+\to K^+ Z^\prime:~
v_\phi \gtrsim ~ 3.0\times 10^7~\text{GeV}\,,\\
 \mu\to eZ^\prime:&~~
v_\phi \gtrsim ~1.3\times 10^7~\text{GeV}\,, \quad
& \tau\to \ell Z^\prime:~
v_\phi \gtrsim~ 7.6\times 10^5~\text{GeV}\,.
\end{align}

\section{Cosmic strings and the gravitational-wave background}
\label{sec:strings}

The previous sections established the model under consideration as well as its implications in the context of flavour physics. We now shift gears and consider the possibility of cosmic strings formation within FN flavour models. More specifically, we are interested in assessing the detectability of the resulting GWB signal from the cosmic strings~\cite{Vachaspati:1984gt,Blanco-Pillado:2011egf,Blanco-Pillado:2013qja,Ringeval:2005kr,Lorenz:2010sm,Gorghetto:2021fsn,Daverio:2015nva,Hindmarsh:2017qff,Baeza-Ballesteros:2023say,Baeza-Ballesteros:2024otj} 
using next-generation GW detectors. The goal is to identify any complementarity of the regions of the parameter space attainable with GW detectors and the flavour searches outlined in the previous sections. We first describe the FN phase transition and the associated formation of cosmic strings. Subsequently, we investigate the dependence of the string width and its tension, two of the key quantities that determine the GWB spectrum, on the fundamental parameters of the model. We then comment on the resulting GWB spectrum induced by the cosmic strings, and conclude by exploring the cosmic string parameter space to assess the detectability of such signals with next-generation GW detectors.

\subsection{The FN phase transition and the formation of cosmic strings}
\label{sec:intro-strings}

In the following, we assume that the reheating temperature after inflation is higher than the FN breaking scale $v_{\phi}$, and hence the FN $U(1)_F$ symmetry is restored at the highest temperature attained in the early Universe.

We do not specify the details of the FN phase transition (PT), simply assuming that at a critical temperature $T_c$
the flavon field $\phi$ undergoes a second order PT developing a vev $v_{\phi}$.
As mentioned, by construction, $\phi$ has $\mathcal{O}(1)$ couplings with the FN messenger fields, which are charged under the SM gauge symmetries and are thus present in the thermal bath down to temperatures of the order of their mass, $\Lambda \sim v_{\phi}$. As a consequence, we expect that thermal corrections induced by FN messenger loops set the FN PT at $T_c \sim v_{\phi}$.

Under the above assumptions, at the temperature $T_c \sim v_{\phi}$ the $U(1)_F$ symmetry is broken and gauge strings are formed in the Universe through the Kibble mechanism.
The string network rapidly approaches the scaling regime where $\mathcal{O}(1)$ strings per Hubble volume are present. A gravitational wave background is then generated by the motion and the contraction of string loops as we will review in the following -- see \cite{Vilenkin:2000jqa} for a review of cosmic strings.

Besides the presence of the cosmic string network, we would like to comment on the expected abundance of the FN sector particles.
The abundance of the FN messenger fields drops when $T$ goes below $\Lambda \sim v_{\phi}$, and then they decay through their mixing with SM fields.
The FN complex scalar is in thermal equilibrium around $T_c$, due to its sizeable couplings to the messengers, hence providing at least a thermal abundance of $Z'$ and of the flavon $\varphi$ (the radial mode of $\phi$) after the PT.
Depending on the gauge coupling, the $Z'$ could be in thermal equilibrium also after $T_c$.
In any case, both $Z'$ and the flavon are produced by the string network and could have a non-negligible abundance in the early Universe.
For these reasons, to be conservative, we will require that both the $Z'$ and the flavon $\varphi$ decay
faster than $0.1$~sec to comply with bounds from Big Bang Nucleosynthesis~(BBN), see e.g.~\cite{Kawasaki:2017bqm}, computing the lifetimes of these particles as discussed in Sections~\ref{sec:flavon} and~\ref{sec:FN-Zprime}.\footnote{We also refer to Ref.~\cite{Lillard:2018zts} where the cosmology of FN models and, in particular, BBN bounds on late-time flavon decays are discussed.}

Finally, we checked that the oscillations of the FN scalar around the minimum do not lead to an early epoch of matter domination -- which may result in a dilution of the GWB signal~\cite{Blasi:2020wpy} -- within the parameter space we are interested in, see Appendix~\ref{app:matter} for details.

\subsection{String profile and its properties}
\label{sec:stringprofileandprop}

The first aspect to investigate is how the shape of the profile, the tension and the width of the string vary as functions of the parameters of the model.
In this subsection, we hence review how to obtain the string solution in our scenario, following \cite{Vilenkin:2000jqa,Hill:1987qx}.

\begin{figure}
\centering
    \includegraphics[width=0.48\textwidth]{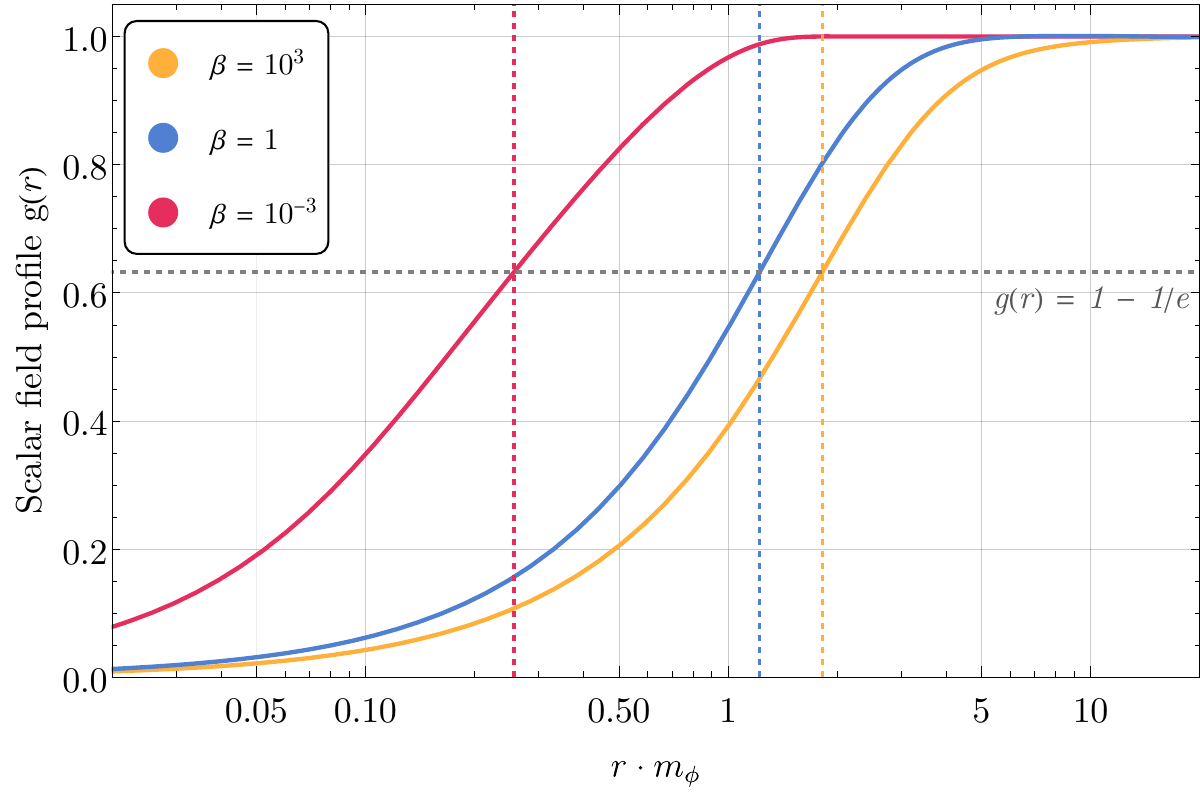}
    \hspace{0.2cm}
    \includegraphics[width=0.48\textwidth]{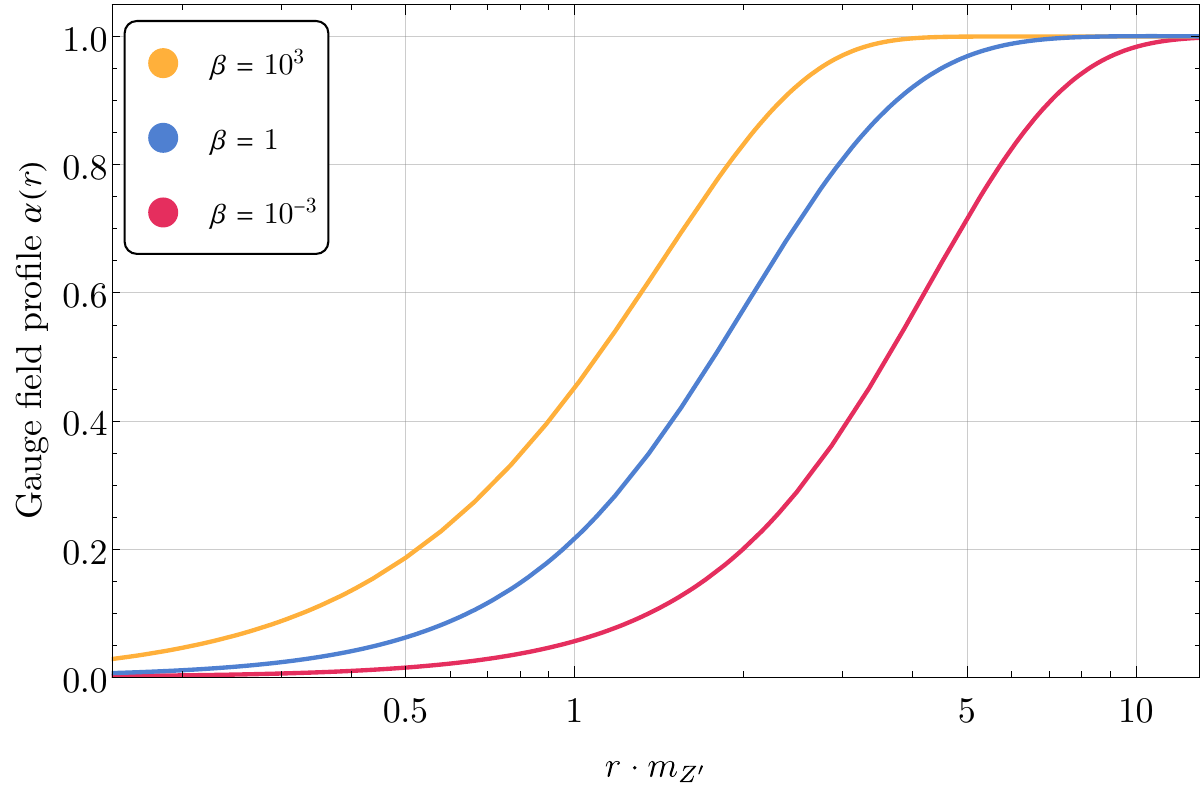}
    \caption{Scalar (left) and gauge (right)  field profiles, $g(r)$ and $\alpha(r)$, respectively, for different values of the $\beta$ parameter, as denoted by the different colours. Each of the field profiles is given as a function of the radius in units of the appropriate mass, i.e.~the scalar mass $m_\varphi$ for the scalar field profile $g(r)$, and the gauge boson mass $m_{Z^\prime}$ for the gauge boson profile $\alpha(r)$. The vertical, dashed lines on the left-hand side denote the intersection of each scalar field profile with the $1-1/e$ line, indicating the width of the string.}     \label{fig:stringprofiles}
\end{figure}

\begin{figure}
\centering
    \includegraphics[width=0.55\textwidth]{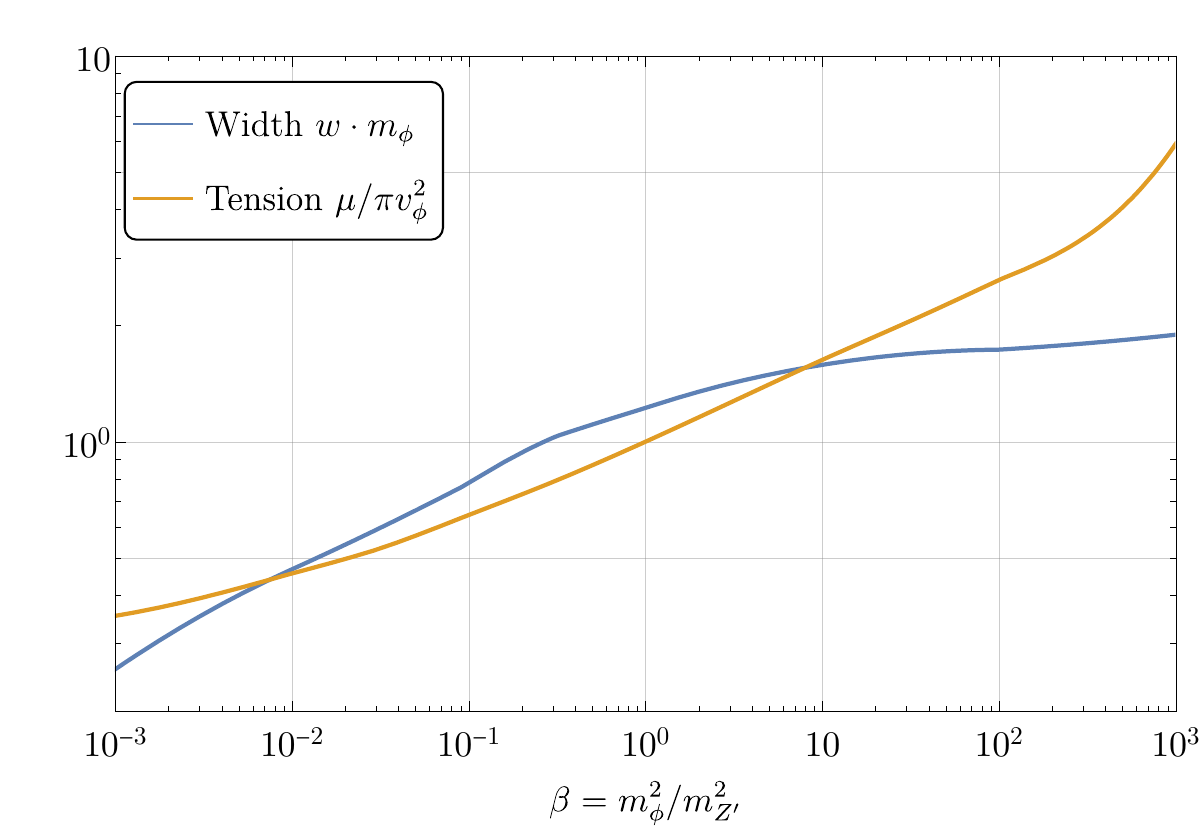}
    \caption{
  Width of the string profile $w$ (blue), as encoded by $W(\beta)$ in Eq.~\eqref{eq:width}, and the string tension $\mu$ (orange), as given by $B(\beta)$ in Eq.~\eqref{eq:tension}, as a function of the mass ratio $\beta = m_{\varphi}^2/m_{Z'}^2$.}
    \label{fig:tensionandwidth}
\end{figure}

We remind the reader that in our $U(1)_F$ model, the flavon $\phi$ obeys the following Lagrangian:
\begin{equation}
\mathcal{L}=
(D^{\mu} \phi)^*
D_{\mu} \phi
-\frac{1}{4}F^\prime_{\mu\nu}F^{\prime\,\mu\nu}-\frac{1}{4}\lambda_\phi\left(|\phi|^2-\eta^2\right)^2,
\end{equation}
where $D_{\mu} = \partial_\mu-ig_F Z^\prime_\mu$, $F^\prime_{\mu\nu}$ is the field strength associated with the gauge field $Z^\prime_{\mu}$, and the parameter $\eta$ is related to the vev defined in Eq.~\eqref{eq:vev} through $\eta=v_\phi/\sqrt{2}$. From this Lagrangian, the following equations of motion can be derived:
\begin{equation}\label{eq:EOM}
   D_{\mu} D^{\mu} 
\phi+\frac{\lambda_\phi}{2}\phi\left(\phi \phi^*-\eta^2\right)=0\,,
\qquad 
    \partial_\mu F^{\prime\,\mu\nu}=
2g_F~\text{Im}\left(\phi^*D^{\nu}\phi\right).
\end{equation}
By adopting the following field and coordinate re-scaling:
\begin{equation}
    \label{eq:rescale}\phi\rightarrow\eta^{-1}\phi\,,~~~~~~~~Z'_\mu\rightarrow\eta^{-1}Z'_\mu\,,~~~~~~~~x_\mu\rightarrow\eta \, g_F \, x_\mu\,,
\end{equation}
the only relevant parameter becomes the squared mass ratio of the flavon and the gauge boson:
\begin{equation}
\label{eq:beta-def}
\beta \equiv \frac{m_\varphi^2}{m_{Z^\prime}^2}= \frac{\lambda_\phi}{2g_F^2}\,.    
\end{equation}
Subsequently, we look for static, cylindrically-symmetric solutions to the equations of motion in Eq.~\eqref{eq:EOM}, i.e.~cosmic strings, of the form:
\begin{equation}
\phi_s(\mathbf{r})=e^{in\theta}g(r)\,,
    \qquad
    Z^\prime_{s,\theta}(\mathbf{r})=-\frac{n}{g_Fr}\alpha(r)\,,
\end{equation}
where the polar coordinates $r^2=x^2+y^2$ and $\theta$ describe the cylindrical string along the $z$ axis, and $n$ is the winding number (taken equal to $1$ in the following). According to the coordinate rescaling in Eq.~\eqref{eq:rescale} and the above ansatz for the field profile, the equations of motion can be rewritten in the following form:
\begin{align}
    & \frac{d^2g}{dr^2}+\frac{1}{r}\frac{dg}{dr}-\frac{n^2 g}{r^2}\left(\alpha-1\right)^2-\beta g\left(g^2-1\right)=0\,, 
    \nonumber
    \\
    & \frac{d^2\alpha}{dr^2}-\frac{1}{r}\frac{d\alpha}{dr}-2g^2\left(\alpha-1\right)=0\,.
    \label{eq:string_eom}
\end{align}
These equations can be solved numerically, as illustrated by the scalar and gauge field profiles provided in Figure~\ref{fig:stringprofiles} for different values of the $\beta$ parameter.
The solution can be used to compute the tension of the string $\mu$:
\begin{equation}
\label{eq:tensionmu}
\mu=\int_0^\infty\int_0^{2\pi}rdrd\theta\left(\left|\frac{\partial\phi}{\partial r}\right|^2+\left|\frac{1}{r}\frac{\partial\phi}{\partial\theta}-ig_FZ^\prime_\theta\phi\right|^2+V(\phi)+\frac{|\mathbf{B}^{\prime}|^2}{2}\right),
\end{equation}
where $\mathbf{B}^{\prime}$ is the magnetic field associated to $U(1)_F$ and hence $|\mathbf{B}^{\prime}|^2= \left| \frac{n}{g_F r}\frac{d\alpha}{dr} \right|^2$.
In practice, it can be shown that this expression reduces to a simple form, which again only depends on the ratio of the masses through $\beta$:
\begin{equation}
\label{eq:tension}
    G\mu=\frac{\pi v_\phi^2}{8\pi M_p^2}B(\beta)
\end{equation}
where $M_p$ is the Planck mass,  $G$ denotes the Newton's constant, $v_\phi$ is the vev defined in Eq.~\eqref{eq:vev}, and $B(\beta)$ is a slowly-varying function of $\beta$. This is illustrated in Figure~\ref{fig:tensionandwidth} with the orange line. Note that constraints from the Cosmic Microwave Background (CMB) limit the maximal value of the string tension to be $G\mu < 10^{-7}$\,\cite{Planck:2015fie,Charnock:2016nzm,Lizarraga:2016onn}.

The numerical solution can also be employed to study the width of the string as a function of the fundamental parameters. This is relevant since, as we will discuss below, a large width can impact the GW spectrum at high frequencies.
To quantify the scaling of the string width $w$ with $\beta$, 
we consider the numerical solution 
for the string profile 
and we
define the width as the coordinate value where the 
scalar field attains the value $(1-1/e)$ times its vev. We then express the width of the string as
\begin{equation}
    \label{eq:width}
    w = \frac{1}{m_{\varphi}} W(\beta),
\end{equation}
where $W(\beta)$ is obtained numerically and shown by the blue line in Figure\,\ref{fig:tensionandwidth}.
As visible also in Figure~\ref{fig:stringprofiles}~(left), the width of the string
scales with $\sim 1/m_{\varphi}$
and can be significantly larger than the tension length scale $1/v_{\phi}$.
%
%
This is also supported by the result for the string width in Figure~\ref{fig:tensionandwidth}, where we note that $w$ indeed approximately scales as $\sim 1/(m_{Z'} \sqrt{\beta}) \sim 1/m_{\varphi}$.

\subsection{Gravitational waves from cosmic strings}
\label{sec:GW_strings}

The strings are expected to generate GWs through various mechanisms, resulting in a gravitational-wave background (GWB). Such a GWB is usually expressed in terms of the dimensionless energy density:
\begin{equation}
    \Omega_\textsc{gw}(f) = \frac{f}{\rho_c}\frac{d\rho_\textsc{gw}}{df},
\end{equation}
where $f$ is the GW frequency and $\rho_c$ is the critical energy density of the Universe. 
To compute the GWB spectrum expected from cosmic strings \cite{Cui_2018,Cui_2019,Gouttenoire:2019kij,Auclair_2020LISA}, one can rely on the velocity-dependent one-scale model, which predicts \cite{Martins:1995tg,Martins_1996,Martins_2002,Sousa_2013,Sousa_2020}:
\begin{equation}
\label{eq:omegaequation}
    \Omega_\textsc{gw}(f)=\sum_{k=1}^\infty\Omega_\textsc{gw}^{(k)}(f)=\frac{8\pi}{3H_0^2}(G\mu)^2f\sum_{k=1}^\infty C_k(f) P_k,
\end{equation}
where $H_0$ is the Hubble rate today, $G$ denotes the Newton's constant, and $\mu$ is the string energy per unit length, i.e.~the tension of the string, as defined in Eq.~\eqref{eq:tension}. The above power spectrum at each frequency $f$ receives contributions from the various harmonics of the string loops, indicated by the index $k$. 
The power corresponding to each harmonic, $P_k$, is given by $P_k=\Gamma/k^q/\zeta(q)$,
for which we assume a cusp--dominated GW emission leading to $q=4/3$.
The normalization constant $\Gamma$ is obtained by matching the total emitted power to the results from numerical simulations, $\Gamma=\sum_k P_k\simeq 50$\,\cite{Blanco-Pillado:2013qja, Blanco_Pillado_2017}.

Additionally, the function $C_k(f)$ is given by
\begin{equation}
    C_k(f)=\frac{2k}{f^2}\int_{t_{\rm scl}}^{t_0}dt \Theta(t)\left(\frac{a(t)}{a(t_0)}\right)^5n(\ell_k,t),
\end{equation}
where $n(\ell_k,t)$ is the number density at the time $t$ of loops with length $l_k$, $a(t)$ is the cosmic scale factor,
and $\Theta(t)$ is a Heaviside function for which the explicit expression is given below.
 The number density evaluates to
\begin{equation}
    n(\ell_k,t)=\frac{F}{t_k^4}\left(\frac{a(t_k)}{a(t)}\right)^3\frac{C_{\rm eff}(t_k)}{\alpha(\alpha+\Gamma G\mu)},
\end{equation}
where $C_{\rm eff}(t_k)$ takes into account whether loops are formed in radiation domination, $C_{\rm eff} \simeq 5.4$, or in matter domination, $C_{\rm eff} \simeq 0.39$, and $F \simeq 0.1$ is an efficiency factor\,\cite{Blanco_Pillado_2014,Sanidas_2012}.
The time $t_k$ corresponds to the formation of the loop whose $k$-th harmonic contributes the present--day frequency $f$, and is given by
\begin{equation}
    t_k=\frac{\ell_k/t+\Gamma G\mu}{\alpha+\Gamma G\mu},~~~~\ell_k=\frac{2k}{f}\frac{a(t)}{a(t_0)}.
\end{equation}
The expression above assumes that, at every time during the evolution of the network, loops are formed with a single length scale given by a constant fraction $\alpha$ of the horizon at formation, with $\alpha \simeq 0.1$ from numerical simulations, and that they shrink only due to energy lost in GWs.

In addition, several Heaviside functions enforce a consistent evolution of the string loops:
\begin{equation}
\label{eq:thetas}
    \Theta(t)=\theta(t_0-t_k)\theta(t_k-t_{\rm scl})\theta(\alpha-\ell_k/t),
\end{equation}
where $t_0$ corresponds to present--today time, and $t_{\rm scl}$ indicates the moment when the string network enters the scaling regime.
Throughout our analysis we will evaluate the scale factor $a(t)$ based on the number of effective degrees of freedom, as given by\,\cite{Saikawa:2020swg}.

The summation over the various harmonics can be simplified by noticing that the $k$-th contribution is related to the fundamental mode as
\begin{equation}
    \Omega_\textsc{gw}^{(k)}(f) = \frac{1}{k^q} \Omega_\textsc{gw}^{(1)}(f/k).
\end{equation}
In addition, for $k \gg 1$, the sum can be more efficiently evaluated by referring to its continuous limit\,\cite{Simone1},
\begin{equation}
    \sum_{k=m}^{k=n}\Omega_\textsc{gw}^{(k)}(f)\approx f^{1-q}\int_{f/n}^{f/m}dx~x^{q-2}\Omega_\textsc{gw}^{(1)}(x).
\end{equation}

However, the basic assumption that the evolution of the string loops is controlled by the energy lost in GWs will eventually break down when the loop reaches a critical size below which particle production becomes the dominant channel.
This critical size, $\ell_c$, has been estimated in\,\cite{Matsunami:2019fss,Auclair:2019jip} for particle emission from cusps to be 
\begin{equation}
\label{eq:cut}
    \ell_c\sim\frac{w}{(\Gamma G\mu)^2},
\end{equation}
where $w$ is the width of the string, and $G\mu$ represents the string tension, as introduced in Eqs.\,\eqref{eq:tension} and \eqref{eq:width}, respectively. Indeed, for lengths $\ell> \ell_c$ the dominant decay mode is identified to be through GW emission, whereas for length scales below $\ell_c$, particle radiation dominates the energy loss.\footnote{Let us stress that the numerical simulations of \,\cite{Hindmarsh:2017qff,Hindmarsh:2021mnl,Baeza-Ballesteros:2024otj} find instead relevant particle production for loops of any size and not only for loops below the critical size in Eq.\,\eqref{eq:cut}. In the following, we shall consider the implications for our flavor model following the findings of Ref.\,\cite{Matsunami:2019fss,Auclair:2019jip}, keeping in mind that our results could change depending on this assumption due to the discrepancy between these results.}
This relation between $\ell_c$ and $w$ has only been verified for $\lambda_\phi, g_F^2 \sim \mathcal{O}(1)$ in the numerical simulation of Ref.~\cite{Matsunami:2019fss}. Nevertheless, we shall assume that this result can be extrapolated to the very small gauge and quartic couplings that we will be interested in. It would be helpful to verify this behavior also for small couplings, $\lambda_\phi, g_F^2 \ll 1$ while keeping their ratio fixed.

Notice that the critical size of the loop in Eq.\,\eqref{eq:cut} is significantly larger than the estimate based on the inverse mass of the emitted particle as taken for instance in Ref.~\cite{Cheng:2024axj}, namely $\ell_c > m_{Z^\prime}^{-1} $, implying a stronger cut for the GW spectrum as long as 
\begin{equation}
\ell_c \, m_{Z^\prime} = \frac{W(\beta)}{\sqrt{\beta} (\Gamma G \mu)^2} \gg 1,
\end{equation} 
which is largely satisfied in our parameter space of interest, $ \beta < 10^3$ and $G\mu < 10^{-7}$, the latter coming from CMB observations as mentioned above.

The effect of the particle radiation cutoff will be visible in the high-frequency part of the GW spectrum, where it will cause the amplitude to decrease above some characteristic frequency associated to the length scale $\ell_c$. This can be taken into account by including an additional Heaviside function in Eq.~\eqref{eq:thetas}:
\begin{equation}
\label{eq:particlecut}
    \Theta(t)=\theta(t_0-t_k)\theta(t_k-t_{\rm scl})\theta(\alpha-\ell_k/t)\theta(\ell_k-\ell_c).
\end{equation}
This corresponds to a high-frequency cutoff in the GW spectrum given by\,\cite{Auclair:2019jip}:
\begin{equation}
    f_{\rm cut} = \left(\frac{8H_0\sqrt{\Omega_R}}{l_c\Gamma G\mu}\right)^{1/2}\simeq 0.0356\left(\frac{G\mu}{10 ^{-10}}\right)^{1/2}\left(\frac{\text{GeV}^{-1}}{w}\right)^{1/2}\rm Hz,
\end{equation}
where $\Omega_R=9.1476\times10^{-5}$ is the radiation density \cite{Planck2018H0}. 

\begin{figure}[t!]
    \centering
    \includegraphics[width=0.75\textwidth]{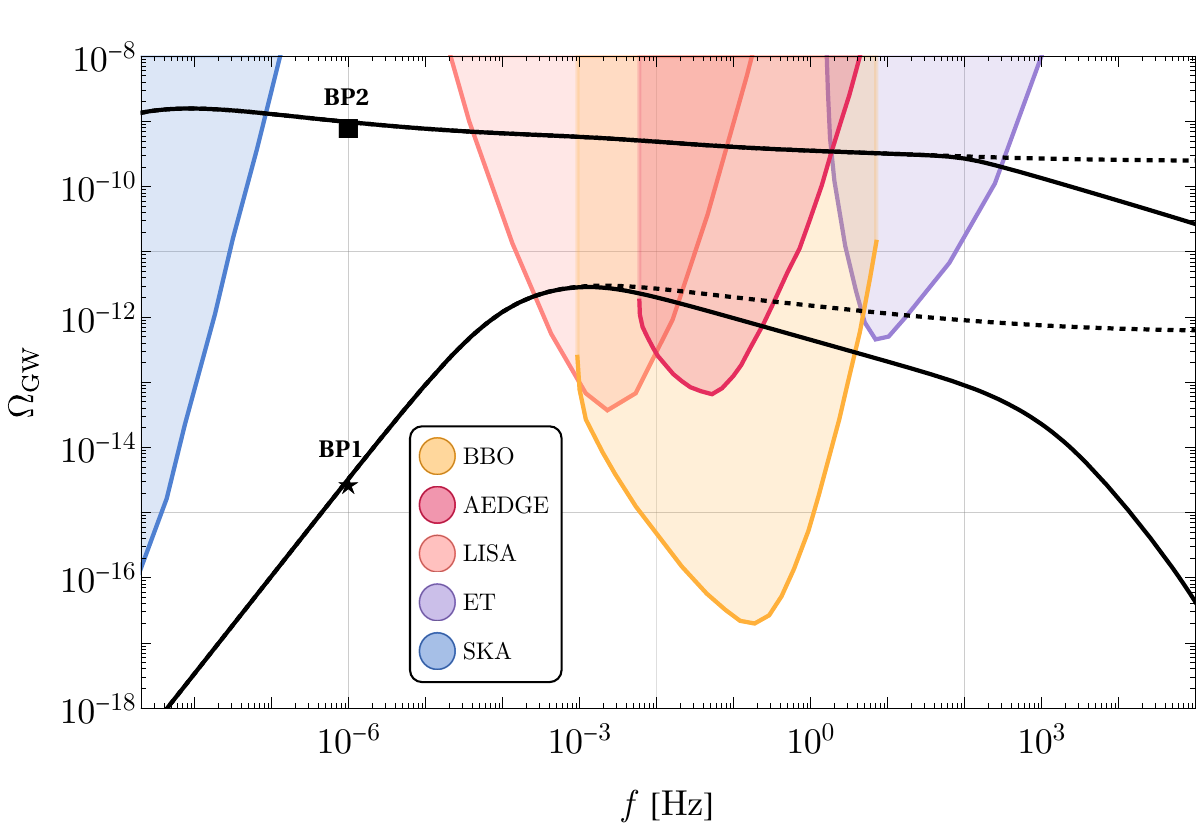}
    \caption{Example of GWB spectra for two benchmark points, labeled BP1 and BP2, as introduced in Eq.~\eqref{eq:benchmarkpoints} and denoted by a square and a star in the parameter space depicted in Figure~\ref{fig:MediumRatio}. The dashed lines denote the GWB spectra computed without taking into account the cut introduced due to the onset of dominant particle radiation, i.e.~as in Eq.~\eqref{eq:thetas}, while the full lines take this cut into account, as captured by Eq.~\eqref{eq:particlecut}. Coloured, shaded regions denote the power-law integrated sensitivity curves for several GW experiments, as obtained from \cite{Schmitz_2021,AEDGE:2019nxb}.}
    \label{fig:examplecutspectrum}
\end{figure}

To illustrate the effect of this cut, we consider two benchmark points,
\begin{equation}\label{eq:benchmarkpoints}
\begin{split}
    &{\rm BP1:}~m_{Z^\prime} = 2 \cdot 10^2~{\rm GeV}, \quad g_F = 10^{-9}, \quad \beta = 1,\\
    &{\rm BP2:}~m_{Z^\prime} = 10^7~{\rm GeV}, \quad g_F = 10^{-7}, \quad \beta=1, 
\end{split}
\end{equation}
in our parameter space of interest and compute the expected GWB spectrum from cosmic strings, both with and without the additional frequency cut due to the onset of dominant particle production. This is shown in Figure~\ref{fig:examplecutspectrum}, which illustrates that the effect of this additional cut can affect the GWB spectrum. However, one notes that depending on the actual value of the frequency cut, the conclusion regarding potential detectability of the GWB signal could remain unaltered, provided the frequency cut happens at large enough frequencies.

\section{Combined GW sensitivity and flavour constraints}
\label{sec:results}

\begin{figure}[t!]
\centering
    \includegraphics[width=0.75\textwidth]{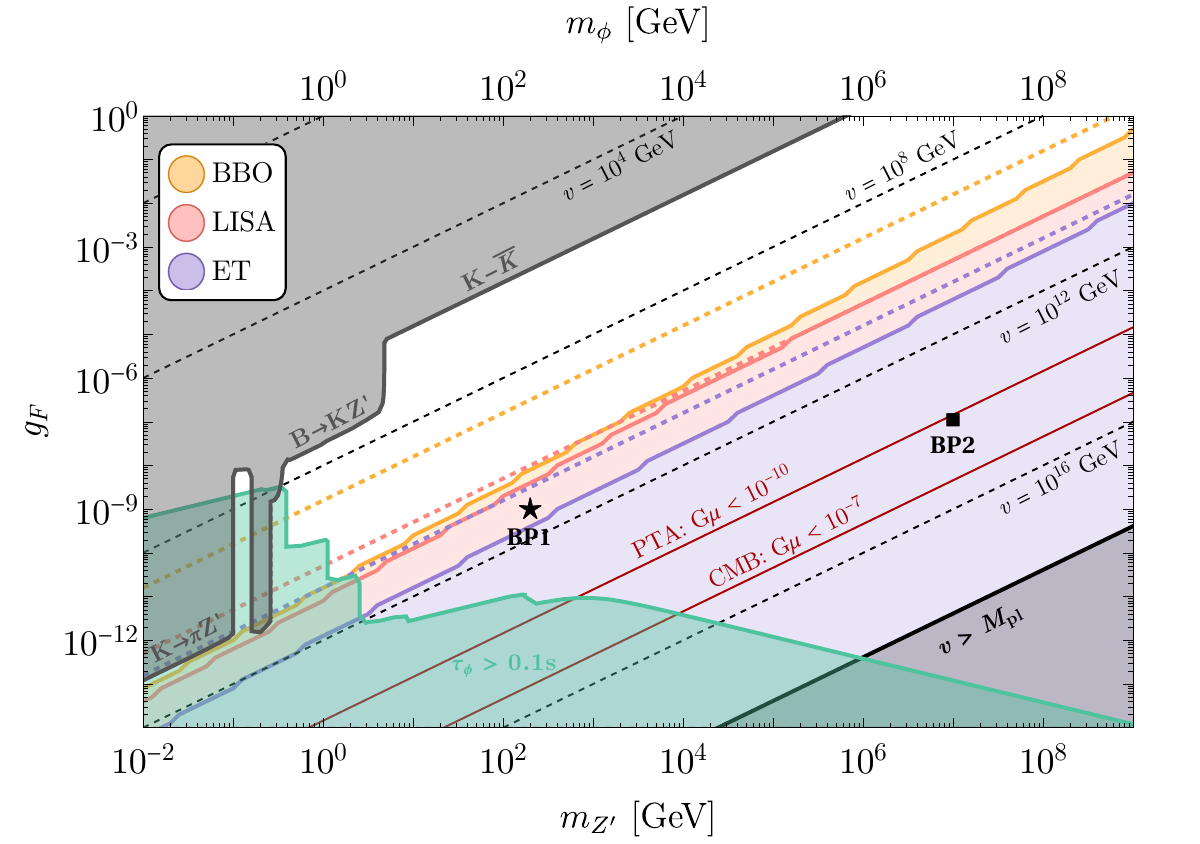}
    \caption{Parameter space scan of $(m_{Z^\prime},g_F)$ for the mass ratio benchmark $\beta=1$. Full lines, together with their coloured regions, denote regions of the parameter space detectable by GW experiments with the onset of dominant particle production taken into account, i.e.~obtained using Eq.~\eqref{eq:particlecut} (BBO in yellow, LISA in pink, and ET in purple). Conversely, the region below the coloured dashed lines indicate the detectability region without accounting for the frequency cut, namely by using Eq.~\eqref{eq:thetas}. The green region denotes part of the parameter space where the lifetime of the flavon $\varphi$ is longer than~0.1~s, potentially interfering with BBN. The grey-shaded region depicts the parameter space excluded by the flavour constraints discussed in Section~\ref{sec:flavour}. The two benchmark points from Eq.~\eqref{eq:benchmarkpoints} are also represented by a star and a square, as in Figure~\ref{fig:examplecutspectrum}.
The regions below the red solid lines are excluded by CMB and PTA data.
}
    \label{fig:MediumRatio}
\end{figure}

\begin{figure}[t!]
    \centering
    \includegraphics[width=0.75\textwidth]{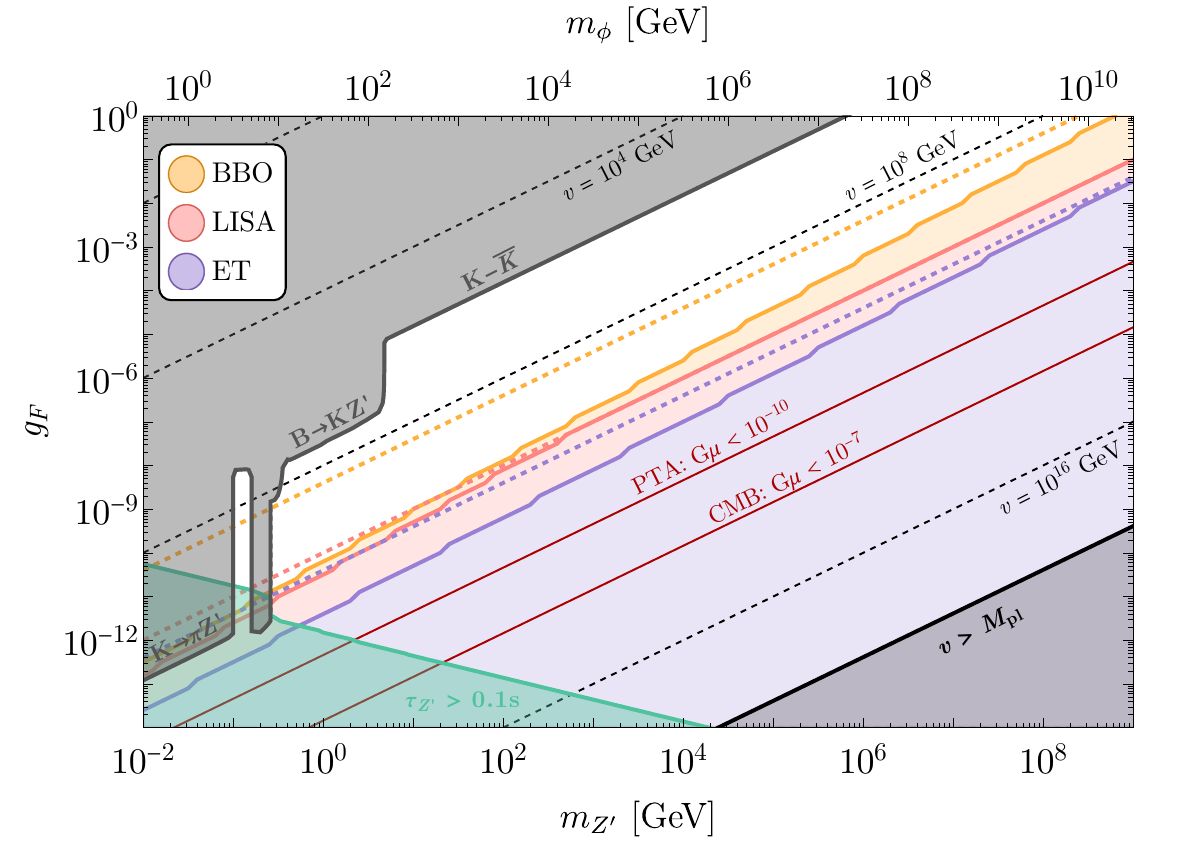}
    \caption{Parameter space scan of $(m_{Z^\prime},g_F)$ for the large mass ratio benchmark where $\beta=10^3$. The green region denotes part of the parameter space where the $Z'$ lifetime is longer than 0.1~s, potentially interfering with BBN. The other coloured regions are as in Figure~\ref{fig:MediumRatio}.}
    \label{fig:LargeRatio}
\end{figure}

We now use the formalism discussed above to explore the detectability of GWB signals originating from cosmic strings using future generation GW experiments.
In this regard, we notice that our type of stable cosmic strings is constrained by current Pulsar Timing Array data, which set a bound on the string tension as stringent as $G\mu < 10^{-11}$\,\cite{NANOGrav:2023hvm}.
We then consider the Einstein Telescope (ET) \cite{Punturo:2010zz,Hild:2010id,Sathyaprakash:2012jk,Maggiore:2019uih}, the Laser Interferometer Space Antenna (LISA) \cite{LISA:2017pwj,Baker:2019nia}, and the Big Bang Observer (BBO) \cite{Corbin:2005ny}, although our results could easily be generalised to other GW experiments.
For each point in the $(m_{Z^\prime},g_F)$ plane, the expected GWB spectrum is computed by means of Eq.~\eqref{eq:omegaequation}, using Eq.~\eqref{eq:particlecut} when the extra cut due to particle production is applied, or using Eq.~\eqref{eq:thetas} when neglecting this additional cut. This spectrum is then compared to the power-law integrated~(PI) sensitivity curve~\cite{Thrane_2013} to obtain an indication of its detectability considering the sensitivities of these future detectors. For each of the experiments, we used the PI curves provided in Ref.~\cite{Schmitz_2021}.

Two different values of the mass ratio $\beta$, defined in Eq.~\eqref{eq:beta-def}, are considered below, corresponding to a case where the gauge boson mass and the scalar mass are of the same order of magnitude (\mbox{$\beta=1$}), and one where the gauge boson is substantially lighter than the scalar (\mbox{$\beta=10^3$}). 
We do not consider the case where the opposite is true, since it would imply some extra tuning in the scalar mass, and also it does not lead to any new phenomenological features.

The detectability region of future GW experiments is illustrated in Figure 
\ref{fig:MediumRatio} and Figure~\ref{fig:LargeRatio} for the $\beta=1$ and $\beta=10^3$ case, respectively,
as a function of the $Z'$ boson mass $m_{Z^\prime}$ and gauge coupling $g_F$.
Both figures show coloured regions corresponding to portions of the parameter space where the expected GWB signal from cosmic strings is large enough to be detectable by the next-generation GW experiments. Note that although not explicitly shown, other experiments such as the Square Kilometer Array (SKA) \cite{Janssen:2014dka} and the AEDGE experiment \cite{AEDGE:2019nxb} display similar detectability regions.

For each case, we consider two possible scenarios. 
Using dashed lines, we show the parameter space that can be probed without implementing the cut on the GW spectrum possibly originating from a large string width
and the resulting particle emission.
Conversely, when we implement the modification to the spectrum due to the string width as explained in Section~\ref{sec:GW_strings}, we illustrate the effect on the detectability of the signal as full lines. In both cases, the  signal associated to any of the parameter space that lies below the coloured line is potentially detectable with the GW experiment under consideration.
As the figures show, the difference between the two scenarios is particularly relevant only in the bottom-left part of the $(m_{Z^\prime},g_F)$ plane, where the combination $m_{Z'} \sqrt{\beta} \sim m_{\varphi}$ is small and hence the width of the strings is large.

In Figure~\ref{fig:MediumRatio}, we also show two benchmark points, defined in Eq.~\eqref{eq:benchmarkpoints}, that illustrate the effect of taking into account the particle emission from strings, which suppresses the GWB signal above some characteristic frequency. These are represented by a square and a star, which refer to the corresponding spectra depicted in Figure~\ref{fig:examplecutspectrum}. In particular, we note that the benchmark point denoted by the star falls within the sensitivity of the ET only if we neglect the frequency cut discussed above, otherwise this part of the parameter space is undetectable. However, the other benchmark point denoted by the square remains detectable, even after the frequency cut is considered. This is also clearly illustrated by the explicit spectra shown in Figure~\ref{fig:examplecutspectrum}.
In general, we see that the modification of the GW spectrum due to large string width effects reduces the detectable portion of the parameter space.

Furthermore, regardless of such effect, by comparing Figure~\ref{fig:MediumRatio} and Figure~\ref{fig:LargeRatio},
we note that the detectable region is slightly larger when $\beta=10^3$ compared to the $\beta=1$ case. This is due to the fact that larger values of $\beta$ result in a larger string tension, and hence the amplitude of the GWB spectrum which is proportional to the string tension increases with $\beta$ as well.

On the same plots, we also add in green the BBN bounds which result from setting $0.1$~s as the upper limit on the lifetime of the $Z'$ and the flavon. These constraints exclude the portion of the plane with very small values of $m_{Z'}$ and $g_F$.
In the $\beta=1$ case, these are dominated by the constraint on the lifetime of the scalar. We note that the shape of the excluded region features some jumps corresponding to the kinematic thresholds of various decay channels. 
In the $\beta=10^3$ case, the BBN bound is dominated instead by the lifetime of the $Z'$, since the flavon is heavier and always decays faster (in particular, the $\varphi\to Z^\prime Z^\prime$ mode is kinematically open).
Note that the threshold features in the case of the flavon are more pronounced, since the scalar couplings with the SM fermions are proportional to the masses of the latter, while the ones of the $Z'$ are proportional to the FN charges and thus have a much milder variation across different fermion species.

Finally, in the figures, we indicate in grey the portion of the parameter space excluded by the combination of the flavour constraints derived in Section~\ref{sec:flavour}.
We note an interesting complementarity between the two types of experimental probes of the model.
Flavour physics observables can probe low to intermediate symmetry-breaking scales, while future GW experiments will test the opposite regime, where the string tension is large enough to imply sizeable GW signals. For instance, even considering the signal deterioration due to particle emission discussed above, we expect that ET can test scales as low as $v_\phi \approx 10^{11}$~GeV and, in the long run, BBO might reach $v_\phi \approx 10^{9}$~GeV.
The combination of the GW and flavour signatures can \emph{close} the parameter space down to a quite restricted allowed portion of the parameter space (the white regions in the figures),  
especially in the light $Z^\prime$ regime, where the flavour constraints become stronger and BBN sets a lower bound on the size of the gauge coupling.
This illustrates the added value of GW searches in the context of flavour physics and shows how particle physics and GW physics can go hand in hand to 
probe the parameter space of models addressing the flavour puzzle.


\section{Summary and outlook}

In this paper we investigated the impact of cosmic strings and the resulting GW signatures within a minimal BSM scenario addressing the flavour puzzle of the SM, based on a local FN-type symmetry.
The formation of the cosmic strings in these types of models is unavoidable, whenever the reheating temperature of the Universe is larger than the FN breaking scale, independently on the nature of the FN phase transition (first or second order),
providing a clear motivation for our investigations.

We first reviewed the FN mechanism, presented a benchmark model, derived the couplings to fermions of the new gauge boson $Z^\prime$ and of the FN Higgs boson (the ``flavon'' $\varphi$), and extensively studied the existing constraints from flavour processes, which enabled us to identify the allowed regions of the parameter space.
Then, we characterised the cosmic strings originating at the FN phase transition, studying  their profile, tension, and width numerically, and we illustrated the features of the resulting GW spectrum.
We also considered the possibility of a string with parametrically large width, and discussed how this affects the GW emission due to particle production, resulting in a drop in the otherwise flat GW spectrum at high frequency.
We then analysed the impact of future GW experiments on the parameter space of the model, showing how GW probes nicely complement flavour physics.
Specifically, we showed that future GW experiments will test scenarios with high symmetry breaking scales, while the flavour experiments constrain low to moderate flavour breaking scales, leaving as undetectable only a range approximately within $10^6\,\text{GeV} \lesssim v_{\phi} \lesssim 10^9\,\text{GeV}$ if the FN gauge boson is heavier than a few GeV. For lighter $Z^\prime$, mesons and leptons can decay into it making the flavour bounds more stringent. Consequently, even without considering limits from BBN, the combination of GW and flavour constraints has the potential to fully test the considered model in the $g_F\ll 1$ regime.

We conclude that, while GW experiments cannot fully \emph{close the gap}, when combined with flavour constraints, they will be able to significantly probe the parameter space of interesting flavour models. 
Furthermore, our results generically highlight the importance of exploring the impact of GW probes into the origin of the SM flavour sector and, in general, particle physics models. 

In the following, we comment on several interesting directions in which our work could be extended in order to further unravel possible connections between flavour models and GW signatures.
In this work, we only considered extending the SM with an abelian flavour symmetry. There are more extended scenarios where the flavour pattern of the SM is explained through the introduction of non-abelian groups, e.g.~$U(2)$~\cite{Linster:2018avp},
and/or where the flavour group is unified with the SM group or an extension of it, see e.g.~\cite{Chen:2023qxi,Chen:2024cht,FernandezNavarro:2023hrf}. 
Also in these contexts, it would be interesting to study the interplay between flavour and GW probes -- see for instance~\cite{Dunsky:2021tih,King:2021gmj,Fu:2023mdu} for recent studies on GW in unifying groups.

Another interesting direction is to consider global FN symmetries. While this reduces the strength of the GW signal from cosmic strings, making it more challenging to detect compared to the gauged scenario\,\cite{Gorghetto:2021fsn,Baeza-Ballesteros:2023say}, it implies the existence of QCD axions in the low energy spectrum,  leading to a solution of the strong CP problem as well as an interesting phenomenology~\cite{Ema:2016ops,Calibbi:2016hwq,Alanne:2018fns}.
In addition, because of the anomalous coupling of the axions with gluons, domain walls (DWs) will be formed at the QCD phase transitions, presenting themselves as another interesting source of a stochastic GWB~\cite{Vilenkin:2000jqa}. 
The DWs must decay to avoid conflict with standard cosmology, and hence in this case a small explicit breaking of the global symmetry must be introduced. This bias sets the strength of the DW signal, while at the same time it might modify the model phenomenology possibly implying further constraints.
We leave the detailed investigation of the global FN symmetry scenario and its interplay with GW signatures for the future.

Finally, during our analysis in Section~\ref{sec:stringprofileandprop}, we illustrated how the large width of the string and the resulting modification of the GW spectrum can impact the reach of GW experiments on relevant portions of the parameter space of the flavour model. 
It would be interesting to extend current numerical simulations 
of the string dynamics to the large width case (namely small couplings) to quantify more precisely the particle emission and the modification of the GW spectrum in these extreme regimes.


\acknowledgments
We would like to thank Ning Chen, Francesco D'Eramo, Tanmay Vachaspati and Robert Ziegler for useful discussions,
and Jiangyi Yi for reporting a mistake in the expressions of the scalar and gauge boson decay widths.
This work is supported by an international cooperation and exchange programme jointly funded by the National Natural Science Foundation of China (NSFC) and the Research Foundation – Flanders (FWO) under the grants No.~12211530479 and VS02223N.
In addition, LC is supported by the NSFC under the grant No.~12035008. 
SB is supported by the Deutsche Forschungsgemeinschaft under Germany’s Excellence Strategy - EXC 2121 Quantum Universe - 390833306, and in part by FWO-Vlaanderen through grant numbers 12B2323N. SB and AM are supported in part by the Strategic Research Program High-Energy Physics of the Research Council of the Vrije Universiteit Brussel.
KT is also supported by the FWO under grant 1179524N.


\appendix

\section{Early matter domination}
\label{app:matter}

Following the analysis in Ref.~\cite{Blasi:2020wpy}, we verified that no early period of matter domination (resulting in a depletion of the GW signal) occurs in our model within the parameters' range of interest.

As discussed in Section~\ref{sec:intro-strings}, we assume that the FN phase transition temperature $T_c$ is of the order of the symmetry breaking vev $v_{\phi}$.
We further assume that the vacuum energy after the FN phase transition redshifts like matter
\begin{equation}
\rho_{\phi} \sim \lambda_\phi v_{\phi}^4 \left(
\frac{T}{T_c}\right)^3\,.
\end{equation}

The above energy density has to be compared with the radiation one, $\rho_{\rm rad} = \frac{\pi^2}{30} g_{*}T^4$, where conservatively we evaluate $g_{*}$ only considering the SM degrees of freedom -- a larger $g_{*}$ would only make the $\rho_{\rm rad}$ relatively more important.
In particular, by equating $\rho_{\phi}$ with $\rho_{\rm rad}$, one can find the temperature, hence the time, where radiation domination would end and an early epoch of matter domination would start, that we denote as $\tau_\textsc{emd}$. 

In order to avoid that to occur, we then require the flavon to decay before the onset of matter domination, that is, $\tau_{\varphi} < \tau_\textsc{emd} $, where the flavon lifetime $\tau_{\varphi}$ can be computed using the formula in Eq.~\eqref{totalwidth_phi}. Notice that $\tau_\varphi$ can be then expressed as a function of the fundamental parameters of our model, $(m_{Z'},\,g_F,\,\beta)$.
For the ranges of $(m_{Z'},g_F)$ and the values of $\beta$ that we adopted in Figures~\ref{fig:MediumRatio} and \ref{fig:LargeRatio}, we find no matter domination in the regions of the parameter space where the BBN bound is satisfied.



\bibliographystyle{JHEP}
\bibliography{biblio.bib}

\providecommand{\href}[2]{#2}\begingroup\raggedright\begin{thebibliography}{100}

\bibitem{Feruglio:2015jfa}
F.~Feruglio, \emph{{Pieces of the Flavour Puzzle}},
  \href{https://doi.org/10.1140/epjc/s10052-015-3576-5}{\emph{Eur. Phys. J. C}
  {\bfseries 75} (2015) 373}
  [\href{https://arxiv.org/abs/1503.04071}{{\ttfamily 1503.04071}}].

\bibitem{Xing:2020ijf}
Z.-z.~Xing, \emph{{Flavor structures of charged fermions and massive
  neutrinos}}, \href{https://doi.org/10.1016/j.physrep.2020.02.001}{\emph{Phys.
  Rept.} {\bfseries 854} (2020) 1}
  [\href{https://arxiv.org/abs/1909.09610}{{\ttfamily 1909.09610}}].

\bibitem{Froggatt:1978nt}
C.D.~Froggatt and H.B.~Nielsen, \emph{{Hierarchy of Quark Masses, Cabibbo
  Angles and CP Violation}},
  \href{https://doi.org/10.1016/0550-3213(79)90316-X}{\emph{Nucl. Phys. B}
  {\bfseries 147} (1979) 277}.

\bibitem{Leurer:1992wg}
M.~Leurer, Y.~Nir and N.~Seiberg, \emph{{Mass matrix models}},
  \href{https://doi.org/10.1016/0550-3213(93)90112-3}{\emph{Nucl. Phys. B}
  {\bfseries 398} (1993) 319}
  [\href{https://arxiv.org/abs/hep-ph/9212278}{{\ttfamily hep-ph/9212278}}].

\bibitem{Leurer:1993gy}
M.~Leurer, Y.~Nir and N.~Seiberg, \emph{{Mass matrix models: The Sequel}},
  \href{https://doi.org/10.1016/0550-3213(94)90074-4}{\emph{Nucl. Phys. B}
  {\bfseries 420} (1994) 468}
  [\href{https://arxiv.org/abs/hep-ph/9310320}{{\ttfamily hep-ph/9310320}}].

\bibitem{Tsumura:2009yf}
K.~Tsumura and L.~Velasco-Sevilla, \emph{{Phenomenology of flavon fields at the
  LHC}}, \href{https://doi.org/10.1103/PhysRevD.81.036012}{\emph{Phys. Rev. D}
  {\bfseries 81} (2010) 036012}
  [\href{https://arxiv.org/abs/0911.2149}{{\ttfamily 0911.2149}}].

\bibitem{Calibbi:2012at}
L.~Calibbi, Z.~Lalak, S.~Pokorski and R.~Ziegler, \emph{{Universal Constraints
  on Low-Energy Flavour Models}},
  \href{https://doi.org/10.1007/JHEP07(2012)004}{\emph{JHEP} {\bfseries 07}
  (2012) 004} [\href{https://arxiv.org/abs/1204.1275}{{\ttfamily 1204.1275}}].

\bibitem{Bauer:2016rxs}
M.~Bauer, T.~Schell and T.~Plehn, \emph{{Hunting the Flavon}},
  \href{https://doi.org/10.1103/PhysRevD.94.056003}{\emph{Phys. Rev. D}
  {\bfseries 94} (2016) 056003}
  [\href{https://arxiv.org/abs/1603.06950}{{\ttfamily 1603.06950}}].

\bibitem{Ema:2016ops}
Y.~Ema, K.~Hamaguchi, T.~Moroi and K.~Nakayama, \emph{{Flaxion: a minimal
  extension to solve puzzles in the standard model}},
  \href{https://doi.org/10.1007/JHEP01(2017)096}{\emph{JHEP} {\bfseries 01}
  (2017) 096} [\href{https://arxiv.org/abs/1612.05492}{{\ttfamily
  1612.05492}}].

\bibitem{Calibbi:2016hwq}
L.~Calibbi, F.~Goertz, D.~Redigolo, R.~Ziegler and J.~Zupan, \emph{{Minimal
  axion model from flavor}},
  \href{https://doi.org/10.1103/PhysRevD.95.095009}{\emph{Phys. Rev. D}
  {\bfseries 95} (2017) 095009}
  [\href{https://arxiv.org/abs/1612.08040}{{\ttfamily 1612.08040}}].

\bibitem{Smolkovic:2019jow}
A.~Smolkovi\v{c}, M.~Tammaro and J.~Zupan, \emph{{Anomaly free Froggatt-Nielsen
  models of flavor}},
  \href{https://doi.org/10.1007/JHEP10(2019)188}{\emph{JHEP} {\bfseries 10}
  (2019) 188} [\href{https://arxiv.org/abs/1907.10063}{{\ttfamily
  1907.10063}}].

\bibitem{Vilenkin:2000jqa}
A.~Vilenkin and E.P.S.~Shellard, \emph{{Cosmic Strings and Other Topological
  Defects}}, Cambridge University Press (7, 2000).

\bibitem{Caldwell:2022qsj}
R.~Caldwell et~al., \emph{{Detection of early-universe gravitational-wave
  signatures and fundamental physics}},
  \href{https://doi.org/10.1007/s10714-022-03027-x}{\emph{Gen. Rel. Grav.}
  {\bfseries 54} (2022) 156}
  [\href{https://arxiv.org/abs/2203.07972}{{\ttfamily 2203.07972}}].

\bibitem{LIGOScientific:2021nrg}
{\scshape LIGO Scientific, Virgo, KAGRA} collaboration, \emph{{Constraints on
  Cosmic Strings Using Data from the Third Advanced LIGO\textendash{}Virgo
  Observing Run}},
  \href{https://doi.org/10.1103/PhysRevLett.126.241102}{\emph{Phys. Rev. Lett.}
  {\bfseries 126} (2021) 241102}
  [\href{https://arxiv.org/abs/2101.12248}{{\ttfamily 2101.12248}}].

\bibitem{NANOGrav:2023gor}
{\scshape NANOGrav} collaboration, \emph{{The NANOGrav 15 yr Data Set: Evidence
  for a Gravitational-wave Background}},
  \href{https://doi.org/10.3847/2041-8213/acdac6}{\emph{Astrophys. J. Lett.}
  {\bfseries 951} (2023) L8}
  [\href{https://arxiv.org/abs/2306.16213}{{\ttfamily 2306.16213}}].

\bibitem{NANOGrav:2023hvm}
{\scshape NANOGrav} collaboration, \emph{{The NANOGrav 15 yr Data Set: Search
  for Signals from New Physics}},
  \href{https://doi.org/10.3847/2041-8213/acdc91}{\emph{Astrophys. J. Lett.}
  {\bfseries 951} (2023) L11}
  [\href{https://arxiv.org/abs/2306.16219}{{\ttfamily 2306.16219}}].

\bibitem{Xu:2023wog}
H.~Xu et~al., \emph{{Searching for the Nano-Hertz Stochastic Gravitational Wave
  Background with the Chinese Pulsar Timing Array Data Release I}},
  \href{https://doi.org/10.1088/1674-4527/acdfa5}{\emph{Res. Astron.
  Astrophys.} {\bfseries 23} (2023) 075024}
  [\href{https://arxiv.org/abs/2306.16216}{{\ttfamily 2306.16216}}].

\bibitem{EPTA:2023fyk}
{\scshape EPTA, InPTA:} collaboration, \emph{{The second data release from the
  European Pulsar Timing Array - III. Search for gravitational wave signals}},
  \href{https://doi.org/10.1051/0004-6361/202346844}{\emph{Astron. Astrophys.}
  {\bfseries 678} (2023) A50}
  [\href{https://arxiv.org/abs/2306.16214}{{\ttfamily 2306.16214}}].

\bibitem{LISACosmologyWorkingGroup:2022jok}
{\scshape LISA Cosmology Working Group} collaboration, \emph{{Cosmology with
  the Laser Interferometer Space Antenna}},
  \href{https://doi.org/10.1007/s41114-023-00045-2}{\emph{Living Rev. Rel.}
  {\bfseries 26} (2023) 5} [\href{https://arxiv.org/abs/2204.05434}{{\ttfamily
  2204.05434}}].

\bibitem{Kibble:1976sj}
T.W.B.~Kibble, \emph{{Topology of Cosmic Domains and Strings}},
  \href{https://doi.org/10.1088/0305-4470/9/8/029}{\emph{J. Phys. A} {\bfseries
  9} (1976) 1387}.

\bibitem{Baldes:2018nel}
I.~Baldes and G.~Servant, \emph{{High scale electroweak phase transition:
  baryogenesis \textbackslash{}\& symmetry non-restoration}},
  \href{https://doi.org/10.1007/JHEP10(2018)053}{\emph{JHEP} {\bfseries 10}
  (2018) 053} [\href{https://arxiv.org/abs/1807.08770}{{\ttfamily
  1807.08770}}].

\bibitem{Baldes:2016gaf}
I.~Baldes, T.~Konstandin and G.~Servant, \emph{{Flavor Cosmology: Dynamical
  Yukawas in the Froggatt-Nielsen Mechanism}},
  \href{https://doi.org/10.1007/JHEP12(2016)073}{\emph{JHEP} {\bfseries 12}
  (2016) 073} [\href{https://arxiv.org/abs/1608.03254}{{\ttfamily
  1608.03254}}].

\bibitem{Greljo:2019xan}
A.~Greljo, T.~Opferkuch and B.A.~Stefanek, \emph{{Gravitational Imprints of
  Flavor Hierarchies}},
  \href{https://doi.org/10.1103/PhysRevLett.124.171802}{\emph{Phys. Rev. Lett.}
  {\bfseries 124} (2020) 171802}
  [\href{https://arxiv.org/abs/1910.02014}{{\ttfamily 1910.02014}}].

\bibitem{Gelmini:2020bqg}
G.B.~Gelmini, S.~Pascoli, E.~Vitagliano and Y.-L.~Zhou, \emph{{Gravitational
  wave signatures from discrete flavor symmetries}},
  \href{https://doi.org/10.1088/1475-7516/2021/02/032}{\emph{JCAP} {\bfseries
  02} (2021) 032} [\href{https://arxiv.org/abs/2009.01903}{{\ttfamily
  2009.01903}}].

\bibitem{Jueid:2023cgp}
A.~Jueid, M.A.~Loualidi, S.~Nasri and M.A.~Ouahid, \emph{{Cosmological domain
  walls from the breaking of S4 flavor symmetry}},
  \href{https://doi.org/10.1103/PhysRevD.109.055048}{\emph{Phys. Rev. D}
  {\bfseries 109} (2024) 055048}
  [\href{https://arxiv.org/abs/2312.04388}{{\ttfamily 2312.04388}}].

\bibitem{Chauhan:2023faf}
G.~Chauhan, P.S.B.~Dev, I.~Dubovyk, B.~Dziewit, W.~Flieger, K.~Grzanka et~al.,
  \emph{{Phenomenology of lepton masses and mixing with discrete flavor
  symmetries}}, \href{https://doi.org/10.1016/j.ppnp.2024.104126}{\emph{Prog.
  Part. Nucl. Phys.} {\bfseries 138} (2024) 104126}
  [\href{https://arxiv.org/abs/2310.20681}{{\ttfamily 2310.20681}}].

\bibitem{Wu:2022tpe}
Y.~Wu, K.-P.~Xie and Y.-L.~Zhou, \emph{{Classification of Abelian domain
  walls}}, \href{https://doi.org/10.1103/PhysRevD.106.075019}{\emph{Phys. Rev.
  D} {\bfseries 106} (2022) 075019}
  [\href{https://arxiv.org/abs/2205.11529}{{\ttfamily 2205.11529}}].

\bibitem{Fu:2024jhu}
B.~Fu, S.F.~King, L.~Marsili, S.~Pascoli, J.~Turner and Y.-L.~Zhou,
  \emph{{Non-Abelian Domain Walls and Gravitational Waves}},
  \href{https://arxiv.org/abs/2409.16359}{{\ttfamily 2409.16359}}.

\bibitem{Ringe:2022rjx}
D.~Ringe, \emph{{Probing intermediate scale Froggatt-Nielsen models at future
  gravitational wave observatories}},
  \href{https://doi.org/10.1103/PhysRevD.107.015030}{\emph{Phys. Rev. D}
  {\bfseries 107} (2023) 015030}
  [\href{https://arxiv.org/abs/2208.07778}{{\ttfamily 2208.07778}}].

\bibitem{Calibbi:2012yj}
L.~Calibbi, Z.~Lalak, S.~Pokorski and R.~Ziegler, \emph{{The Messenger Sector
  of SUSY Flavour Models and Radiative Breaking of Flavour Universality}},
  \href{https://doi.org/10.1007/JHEP06(2012)018}{\emph{JHEP} {\bfseries 06}
  (2012) 018} [\href{https://arxiv.org/abs/1203.1489}{{\ttfamily 1203.1489}}].

\bibitem{Fedele:2020fvh}
M.~Fedele, A.~Mastroddi and M.~Valli, \emph{{Minimal Froggatt-Nielsen
  textures}}, \href{https://doi.org/10.1007/JHEP03(2021)135}{\emph{JHEP}
  {\bfseries 03} (2021) 135}
  [\href{https://arxiv.org/abs/2009.05587}{{\ttfamily 2009.05587}}].

\bibitem{Nishimura:2020nre}
S.~Nishimura, C.~Miyao and H.~Otsuka, \emph{{Exploring the flavor structure of
  quarks and leptons with reinforcement learning}},
  \href{https://doi.org/10.1007/JHEP12(2023)021}{\emph{JHEP} {\bfseries 23}
  (2020) 021} [\href{https://arxiv.org/abs/2304.14176}{{\ttfamily
  2304.14176}}].

\bibitem{Cornella:2023zme}
C.~Cornella, D.~Curtin, E.T.~Neil and J.O.~Thompson, \emph{{Mapping and Probing
  Froggatt-Nielsen Solutions to the Quark Flavor Puzzle}},
  \href{https://arxiv.org/abs/2306.08026}{{\ttfamily 2306.08026}}.

\bibitem{Weinberg:1979sa}
S.~Weinberg, \emph{{Baryon and Lepton Nonconserving Processes}},
  \href{https://doi.org/10.1103/PhysRevLett.43.1566}{\emph{Phys. Rev. Lett.}
  {\bfseries 43} (1979) 1566}.

\bibitem{Hall:1999sn}
L.J.~Hall, H.~Murayama and N.~Weiner, \emph{{Neutrino mass anarchy}},
  \href{https://doi.org/10.1103/PhysRevLett.84.2572}{\emph{Phys. Rev. Lett.}
  {\bfseries 84} (2000) 2572}
  [\href{https://arxiv.org/abs/hep-ph/9911341}{{\ttfamily hep-ph/9911341}}].

\bibitem{Esteban:2020cvm}
I.~Esteban, M.C.~Gonzalez-Garcia, M.~Maltoni, T.~Schwetz and A.~Zhou,
  \emph{{The fate of hints: updated global analysis of three-flavor neutrino
  oscillations}}, \href{https://doi.org/10.1007/JHEP09(2020)178}{\emph{JHEP}
  {\bfseries 09} (2020) 178}
  [\href{https://arxiv.org/abs/2007.14792}{{\ttfamily 2007.14792}}].

\bibitem{Altarelli:2012ia}
G.~Altarelli, F.~Feruglio, I.~Masina and L.~Merlo, \emph{{Repressing Anarchy in
  Neutrino Mass Textures}},
  \href{https://doi.org/10.1007/JHEP11(2012)139}{\emph{JHEP} {\bfseries 11}
  (2012) 139} [\href{https://arxiv.org/abs/1207.0587}{{\ttfamily 1207.0587}}].

\bibitem{Bergstrom:2014owa}
J.~Bergstrom, D.~Meloni and L.~Merlo, \emph{{Bayesian comparison of U(1) lepton
  flavor models}},
  \href{https://doi.org/10.1103/PhysRevD.89.093021}{\emph{Phys. Rev. D}
  {\bfseries 89} (2014) 093021}
  [\href{https://arxiv.org/abs/1403.4528}{{\ttfamily 1403.4528}}].

\bibitem{Davidson:1981zd}
A.~Davidson and K.C.~Wali, \emph{{Minimal Flavor Unification via
  Multigenerational Peccei-Quinn Symmetry}},
  \href{https://doi.org/10.1103/PhysRevLett.48.11}{\emph{Phys. Rev. Lett.}
  {\bfseries 48} (1982) 11}.

\bibitem{Wilczek:1982rv}
F.~Wilczek, \emph{{Axions and Family Symmetry Breaking}},
  \href{https://doi.org/10.1103/PhysRevLett.49.1549}{\emph{Phys. Rev. Lett.}
  {\bfseries 49} (1982) 1549}.

\bibitem{Reiss:1982sq}
D.B.~Reiss, \emph{{Can the Family Group Be a Global Symmetry?}},
  \href{https://doi.org/10.1016/0370-2693(82)90647-5}{\emph{Phys. Lett. B}
  {\bfseries 115} (1982) 217}.

\bibitem{Davidson:1983fy}
A.~Davidson, V.P.~Nair and K.C.~Wali, \emph{{Peccei-Quinn Symmetry as Flavor
  Symmetry and Grand Unification}},
  \href{https://doi.org/10.1103/PhysRevD.29.1504}{\emph{Phys. Rev. D}
  {\bfseries 29} (1984) 1504}.

\bibitem{Chang:1987hz}
D.~Chang and G.~Senjanovic, \emph{{On Axion and Familons}},
  \href{https://doi.org/10.1016/0370-2693(87)90012-8}{\emph{Phys. Lett. B}
  {\bfseries 188} (1987) 231}.

\bibitem{Berezhiani:1990jj}
Z.G.~Berezhiani and M.Y.~Khlopov, \emph{{Physical and astrophysical
  consequences of breaking of the symmetry of families. (In Russian)}},
  {\emph{Sov. J. Nucl. Phys.} {\bfseries 51} (1990) 935}.

\bibitem{Alonso:2018bcg}
R.~Alonso, A.~Carmona, B.M.~Dillon, J.F.~Kamenik, J.~Martin~Camalich and
  J.~Zupan, \emph{{A clockwork solution to the flavor puzzle}},
  \href{https://doi.org/10.1007/JHEP10(2018)099}{\emph{JHEP} {\bfseries 10}
  (2018) 099} [\href{https://arxiv.org/abs/1807.09792}{{\ttfamily
  1807.09792}}].

\bibitem{Bonnefoy:2019lsn}
Q.~Bonnefoy, E.~Dudas and S.~Pokorski, \emph{{Chiral Froggatt-Nielsen models,
  gauge anomalies and flavourful axions}},
  \href{https://doi.org/10.1007/JHEP01(2020)191}{\emph{JHEP} {\bfseries 01}
  (2020) 191} [\href{https://arxiv.org/abs/1909.05336}{{\ttfamily
  1909.05336}}].

\bibitem{Kang:2004bz}
J.~Kang and P.~Langacker, \emph{{$Z^\prime$ discovery limits for supersymmetric
  E(6) models}}, \href{https://doi.org/10.1103/PhysRevD.71.035014}{\emph{Phys.
  Rev. D} {\bfseries 71} (2005) 035014}
  [\href{https://arxiv.org/abs/hep-ph/0412190}{{\ttfamily hep-ph/0412190}}].

\bibitem{Landau:1948kw}
L.D.~Landau, \emph{{On the angular momentum of a system of two photons}},
  \href{https://doi.org/10.1016/B978-0-08-010586-4.50070-5}{\emph{Dokl. Akad.
  Nauk SSSR} {\bfseries 60} (1948) 207}.

\bibitem{Yang:1950rg}
C.-N.~Yang, \emph{{Selection Rules for the Dematerialization of a Particle Into
  Two Photons}}, \href{https://doi.org/10.1103/PhysRev.77.242}{\emph{Phys.
  Rev.} {\bfseries 77} (1950) 242}.

\bibitem{Redondo:2008ec}
J.~Redondo and M.~Postma, \emph{{Massive hidden photons as lukewarm dark
  matter}}, \href{https://doi.org/10.1088/1475-7516/2009/02/005}{\emph{JCAP}
  {\bfseries 02} (2009) 005} [\href{https://arxiv.org/abs/0811.0326}{{\ttfamily
  0811.0326}}].

\bibitem{Calibbi:2015sfa}
L.~Calibbi, A.~Crivellin and B.~Zald\'\i{}var, \emph{{Flavor portal to dark
  matter}}, \href{https://doi.org/10.1103/PhysRevD.92.016004}{\emph{Phys. Rev.
  D} {\bfseries 92} (2015) 016004}
  [\href{https://arxiv.org/abs/1501.07268}{{\ttfamily 1501.07268}}].

\bibitem{MartinCamalich:2020dfe}
J.~Martin~Camalich, M.~Pospelov, P.N.H.~Vuong, R.~Ziegler and J.~Zupan,
  \emph{{Quark Flavor Phenomenology of the QCD Axion}},
  \href{https://doi.org/10.1103/PhysRevD.102.015023}{\emph{Phys. Rev. D}
  {\bfseries 102} (2020) 015023}
  [\href{https://arxiv.org/abs/2002.04623}{{\ttfamily 2002.04623}}].

\bibitem{Silvestrini:2018dos}
L.~Silvestrini and M.~Valli, \emph{{Model-independent Bounds on the Standard
  Model Effective Theory from Flavour Physics}},
  \href{https://doi.org/10.1016/j.physletb.2019.135062}{\emph{Phys. Lett. B}
  {\bfseries 799} (2019) 135062}
  [\href{https://arxiv.org/abs/1812.10913}{{\ttfamily 1812.10913}}].

\bibitem{Calibbi:2017uvl}
L.~Calibbi and G.~Signorelli, \emph{{Charged Lepton Flavour Violation: An
  Experimental and Theoretical Introduction}},
  \href{https://doi.org/10.1393/ncr/i2018-10144-0}{\emph{Riv. Nuovo Cim.}
  {\bfseries 41} (2018) 71} [\href{https://arxiv.org/abs/1709.00294}{{\ttfamily
  1709.00294}}].

\bibitem{SINDRUM:1987nra}
{\scshape SINDRUM} collaboration, \emph{{Search for the Decay $\mu^+ \to e^+
  e^+ e^-$}}, \href{https://doi.org/10.1016/0550-3213(88)90462-2}{\emph{Nucl.
  Phys. B} {\bfseries 299} (1988) 1}.

\bibitem{SINDRUMII:2006dvw}
{\scshape SINDRUM II} collaboration, \emph{{A Search for muon to electron
  conversion in muonic gold}},
  \href{https://doi.org/10.1140/epjc/s2006-02582-x}{\emph{Eur. Phys. J. C}
  {\bfseries 47} (2006) 337}.

\bibitem{Altmannshofer:2023hkn}
W.~Altmannshofer, A.~Crivellin, H.~Haigh, G.~Inguglia and J.~Martin~Camalich,
  \emph{{Light new physics in $B\to K^{(*)}\nu\bar\nu$?}},
  \href{https://doi.org/10.1103/PhysRevD.109.075008}{\emph{Phys. Rev. D}
  {\bfseries 109} (2024) 075008}
  [\href{https://arxiv.org/abs/2311.14629}{{\ttfamily 2311.14629}}].

\bibitem{PDG}
{\scshape Particle Data Group} collaboration, \emph{{Review of Particle
  Physics}}, \href{https://doi.org/10.1093/ptep/ptac097}{\emph{PTEP} {\bfseries
  2022} (2022) 083C01}.

\bibitem{Carrasco:2016kpy}
N.~Carrasco, P.~Lami, V.~Lubicz, L.~Riggio, S.~Simula and C.~Tarantino,
  \emph{{$K \to \pi$ semileptonic form factors with $N_f=2+1+1$ twisted mass
  fermions}}, \href{https://doi.org/10.1103/PhysRevD.93.114512}{\emph{Phys.
  Rev. D} {\bfseries 93} (2016) 114512}
  [\href{https://arxiv.org/abs/1602.04113}{{\ttfamily 1602.04113}}].

\bibitem{Gubernari:2018wyi}
N.~Gubernari, A.~Kokulu and D.~van Dyk, \emph{{$B\to P$ and $B\to V$ Form
  Factors from $B$-Meson Light-Cone Sum Rules beyond Leading Twist}},
  \href{https://doi.org/10.1007/JHEP01(2019)150}{\emph{JHEP} {\bfseries 01}
  (2019) 150} [\href{https://arxiv.org/abs/1811.00983}{{\ttfamily
  1811.00983}}].

\bibitem{Heeck:2016xkh}
J.~Heeck, \emph{{Lepton flavor violation with light vector bosons}},
  \href{https://doi.org/10.1016/j.physletb.2016.05.007}{\emph{Phys. Lett. B}
  {\bfseries 758} (2016) 101}
  [\href{https://arxiv.org/abs/1602.03810}{{\ttfamily 1602.03810}}].

\bibitem{Ibarra:2021xyk}
A.~Ibarra, M.~Mar\'\i{}n and P.~Roig, \emph{{Flavor violating muon decay into
  an electron and a light gauge boson}},
  \href{https://doi.org/10.1016/j.physletb.2022.136933}{\emph{Phys. Lett. B}
  {\bfseries 827} (2022) 136933}
  [\href{https://arxiv.org/abs/2110.03737}{{\ttfamily 2110.03737}}].

\bibitem{NA62:2021zjw}
{\scshape NA62} collaboration, \emph{{Measurement of the very rare K$^{+}\to
  {\pi}^{+}\nu \overline{\nu} $ decay}},
  \href{https://doi.org/10.1007/JHEP06(2021)093}{\emph{JHEP} {\bfseries 06}
  (2021) 093} [\href{https://arxiv.org/abs/2103.15389}{{\ttfamily
  2103.15389}}].

\bibitem{Belle-II:2022heu}
{\scshape Belle-II} collaboration, \emph{{Search for Lepton-Flavor-Violating
  \ensuremath{\tau} Decays to a Lepton and an Invisible Boson at Belle II}},
  \href{https://doi.org/10.1103/PhysRevLett.130.181803}{\emph{Phys. Rev. Lett.}
  {\bfseries 130} (2023) 181803}
  [\href{https://arxiv.org/abs/2212.03634}{{\ttfamily 2212.03634}}].

\bibitem{Jodidio:1986mz}
A.~Jodidio et~al., \emph{{Search for Right-Handed Currents in Muon Decay}},
  \href{https://doi.org/10.1103/PhysRevD.34.1967}{\emph{Phys. Rev. D}
  {\bfseries 34} (1986) 1967}.

\bibitem{Calibbi:2020jvd}
L.~Calibbi, D.~Redigolo, R.~Ziegler and J.~Zupan, \emph{{Looking forward to
  lepton-flavor-violating ALPs}},
  \href{https://doi.org/10.1007/JHEP09(2021)173}{\emph{JHEP} {\bfseries 09}
  (2021) 173} [\href{https://arxiv.org/abs/2006.04795}{{\ttfamily
  2006.04795}}].

\bibitem{TWIST:2014ymv}
{\scshape TWIST} collaboration, \emph{{Search for two body muon decay
  signals}}, \href{https://doi.org/10.1103/PhysRevD.91.052020}{\emph{Phys. Rev.
  D} {\bfseries 91} (2015) 052020}
  [\href{https://arxiv.org/abs/1409.0638}{{\ttfamily 1409.0638}}].

\bibitem{Vachaspati:1984gt}
T.~Vachaspati and A.~Vilenkin, \emph{{Gravitational Radiation from Cosmic
  Strings}}, \href{https://doi.org/10.1103/PhysRevD.31.3052}{\emph{Phys. Rev.
  D} {\bfseries 31} (1985) 3052}.

\bibitem{Blanco-Pillado:2011egf}
J.J.~Blanco-Pillado, K.D.~Olum and B.~Shlaer, \emph{{Large parallel cosmic
  string simulations: New results on loop production}},
  \href{https://doi.org/10.1103/PhysRevD.83.083514}{\emph{Phys. Rev. D}
  {\bfseries 83} (2011) 083514}
  [\href{https://arxiv.org/abs/1101.5173}{{\ttfamily 1101.5173}}].

\bibitem{Blanco-Pillado:2013qja}
J.J.~Blanco-Pillado, K.D.~Olum and B.~Shlaer, \emph{{The number of cosmic
  string loops}}, \href{https://doi.org/10.1103/PhysRevD.89.023512}{\emph{Phys.
  Rev. D} {\bfseries 89} (2014) 023512}
  [\href{https://arxiv.org/abs/1309.6637}{{\ttfamily 1309.6637}}].

\bibitem{Ringeval:2005kr}
C.~Ringeval, M.~Sakellariadou and F.~Bouchet, \emph{{Cosmological evolution of
  cosmic string loops}},
  \href{https://doi.org/10.1088/1475-7516/2007/02/023}{\emph{JCAP} {\bfseries
  02} (2007) 023} [\href{https://arxiv.org/abs/astro-ph/0511646}{{\ttfamily
  astro-ph/0511646}}].

\bibitem{Lorenz:2010sm}
L.~Lorenz, C.~Ringeval and M.~Sakellariadou, \emph{{Cosmic string loop
  distribution on all length scales and at any redshift}},
  \href{https://doi.org/10.1088/1475-7516/2010/10/003}{\emph{JCAP} {\bfseries
  10} (2010) 003} [\href{https://arxiv.org/abs/1006.0931}{{\ttfamily
  1006.0931}}].

\bibitem{Gorghetto:2021fsn}
M.~Gorghetto, E.~Hardy and H.~Nicolaescu, \emph{{Observing invisible axions
  with gravitational waves}},
  \href{https://doi.org/10.1088/1475-7516/2021/06/034}{\emph{JCAP} {\bfseries
  06} (2021) 034} [\href{https://arxiv.org/abs/2101.11007}{{\ttfamily
  2101.11007}}].

\bibitem{Daverio:2015nva}
D.~Daverio, M.~Hindmarsh, M.~Kunz, J.~Lizarraga and J.~Urrestilla,
  \emph{{Energy-momentum correlations for Abelian Higgs cosmic strings}},
  \href{https://doi.org/10.1103/PhysRevD.95.049903}{\emph{Phys. Rev. D}
  {\bfseries 93} (2016) 085014}
  [\href{https://arxiv.org/abs/1510.05006}{{\ttfamily 1510.05006}}].

\bibitem{Hindmarsh:2017qff}
M.~Hindmarsh, J.~Lizarraga, J.~Urrestilla, D.~Daverio and M.~Kunz,
  \emph{{Scaling from gauge and scalar radiation in Abelian Higgs string
  networks}}, \href{https://doi.org/10.1103/PhysRevD.96.023525}{\emph{Phys.
  Rev. D} {\bfseries 96} (2017) 023525}
  [\href{https://arxiv.org/abs/1703.06696}{{\ttfamily 1703.06696}}].

\bibitem{Baeza-Ballesteros:2023say}
J.~Baeza-Ballesteros, E.J.~Copeland, D.G.~Figueroa and J.~Lizarraga,
  \emph{{Gravitational wave emission from a cosmic string loop: Global case}},
  \href{https://doi.org/10.1103/PhysRevD.110.043522}{\emph{Phys. Rev. D}
  {\bfseries 110} (2024) 043522}
  [\href{https://arxiv.org/abs/2308.08456}{{\ttfamily 2308.08456}}].

\bibitem{Baeza-Ballesteros:2024otj}
J.~Baeza-Ballesteros, E.J.~Copeland, D.G.~Figueroa and J.~Lizarraga,
  \emph{{Gravitational Wave Emission from Cosmic String Loops, II: Local
  Case}},  \href{https://arxiv.org/abs/2408.02364}{{\ttfamily 2408.02364}}.

\bibitem{Kawasaki:2017bqm}
M.~Kawasaki, K.~Kohri, T.~Moroi and Y.~Takaesu, \emph{{Revisiting Big-Bang
  Nucleosynthesis Constraints on Long-Lived Decaying Particles}},
  \href{https://doi.org/10.1103/PhysRevD.97.023502}{\emph{Phys. Rev. D}
  {\bfseries 97} (2018) 023502}
  [\href{https://arxiv.org/abs/1709.01211}{{\ttfamily 1709.01211}}].

\bibitem{Lillard:2018zts}
B.~Lillard, M.~Ratz, T.~Tait, M.~P. and S.~Trojanowski, \emph{{The Flavor of
  Cosmology}}, \href{https://doi.org/10.1088/1475-7516/2018/07/056}{\emph{JCAP}
  {\bfseries 07} (2018) 056}
  [\href{https://arxiv.org/abs/1804.03662}{{\ttfamily 1804.03662}}].

\bibitem{Blasi:2020wpy}
S.~Blasi, V.~Brdar and K.~Schmitz, \emph{{Fingerprint of low-scale leptogenesis
  in the primordial gravitational-wave spectrum}},
  \href{https://doi.org/10.1103/PhysRevResearch.2.043321}{\emph{Phys. Rev.
  Res.} {\bfseries 2} (2020) 043321}
  [\href{https://arxiv.org/abs/2004.02889}{{\ttfamily 2004.02889}}].

\bibitem{Hill:1987qx}
C.T.~Hill, H.M.~Hodges and M.S.~Turner, \emph{{Bosonic Superconducting Cosmic
  Strings}}, \href{https://doi.org/10.1103/PhysRevD.37.263}{\emph{Phys. Rev. D}
  {\bfseries 37} (1988) 263}.

\bibitem{Planck:2015fie}
{\scshape Planck} collaboration, \emph{{Planck 2015 results. XIII. Cosmological
  parameters}},
  \href{https://doi.org/10.1051/0004-6361/201525830}{\emph{Astron. Astrophys.}
  {\bfseries 594} (2016) A13}
  [\href{https://arxiv.org/abs/1502.01589}{{\ttfamily 1502.01589}}].

\bibitem{Charnock:2016nzm}
T.~Charnock, A.~Avgoustidis, E.J.~Copeland and A.~Moss, \emph{{CMB constraints
  on cosmic strings and superstrings}},
  \href{https://doi.org/10.1103/PhysRevD.93.123503}{\emph{Phys. Rev. D}
  {\bfseries 93} (2016) 123503}
  [\href{https://arxiv.org/abs/1603.01275}{{\ttfamily 1603.01275}}].

\bibitem{Lizarraga:2016onn}
J.~Lizarraga, J.~Urrestilla, D.~Daverio, M.~Hindmarsh and M.~Kunz, \emph{{New
  CMB constraints for Abelian Higgs cosmic strings}},
  \href{https://doi.org/10.1088/1475-7516/2016/10/042}{\emph{JCAP} {\bfseries
  10} (2016) 042} [\href{https://arxiv.org/abs/1609.03386}{{\ttfamily
  1609.03386}}].

\bibitem{Cui_2018}
Y.~Cui, M.~Lewicki, D.E.~Morrissey and J.D.~Wells, \emph{{Cosmic Archaeology
  with Gravitational Waves from Cosmic Strings}},
  \href{https://doi.org/10.1103/PhysRevD.97.123505}{\emph{Phys. Rev. D}
  {\bfseries 97} (2018) 123505}
  [\href{https://arxiv.org/abs/1711.03104}{{\ttfamily 1711.03104}}].

\bibitem{Cui_2019}
Y.~Cui, M.~Lewicki, D.E.~Morrissey and J.D.~Wells, \emph{{Probing the pre-BBN
  universe with gravitational waves from cosmic strings}},
  \href{https://doi.org/10.1007/JHEP01(2019)081}{\emph{JHEP} {\bfseries 01}
  (2019) 081} [\href{https://arxiv.org/abs/1808.08968}{{\ttfamily
  1808.08968}}].

\bibitem{Gouttenoire:2019kij}
Y.~Gouttenoire, G.~Servant and P.~Simakachorn, \emph{{Beyond the Standard
  Models with Cosmic Strings}},
  \href{https://doi.org/10.1088/1475-7516/2020/07/032}{\emph{JCAP} {\bfseries
  07} (2020) 032} [\href{https://arxiv.org/abs/1912.02569}{{\ttfamily
  1912.02569}}].

\bibitem{Auclair_2020LISA}
P.~Auclair, J.J.~Blanco-Pillado, D.G.~Figueroa, A.C.~Jenkins, M.~Lewicki,
  M.~Sakellariadou et~al., \emph{Probing the gravitational wave background from
  cosmic strings with lisa},
  \href{https://doi.org/10.1088/1475-7516/2020/04/034}{\emph{Journal of
  Cosmology and Astroparticle Physics} {\bfseries 2020} (2020) 034–034}.

\bibitem{Martins:1995tg}
C.J.A.P.~Martins and E.P.S.~Shellard, \emph{{String evolution with friction}},
  \href{https://doi.org/10.1103/PhysRevD.53.R575}{\emph{Phys. Rev. D}
  {\bfseries 53} (1996) 575}
  [\href{https://arxiv.org/abs/hep-ph/9507335}{{\ttfamily hep-ph/9507335}}].

\bibitem{Martins_1996}
C.J.A.P.~Martins and E.P.S.~Shellard, \emph{{Quantitative string evolution}},
  \href{https://doi.org/10.1103/PhysRevD.54.2535}{\emph{Phys. Rev. D}
  {\bfseries 54} (1996) 2535}
  [\href{https://arxiv.org/abs/hep-ph/9602271}{{\ttfamily hep-ph/9602271}}].

\bibitem{Martins_2002}
C.J.A.P.~Martins and E.P.S.~Shellard, \emph{{Extending the velocity dependent
  one scale string evolution model}},
  \href{https://doi.org/10.1103/PhysRevD.65.043514}{\emph{Phys. Rev. D}
  {\bfseries 65} (2002) 043514}
  [\href{https://arxiv.org/abs/hep-ph/0003298}{{\ttfamily hep-ph/0003298}}].

\bibitem{Sousa_2013}
L.~Sousa and P.P.~Avelino, \emph{{Stochastic Gravitational Wave Background
  generated by Cosmic String Networks: Velocity-Dependent One-Scale model
  versus Scale-Invariant Evolution}},
  \href{https://doi.org/10.1103/PhysRevD.88.023516}{\emph{Phys. Rev. D}
  {\bfseries 88} (2013) 023516}
  [\href{https://arxiv.org/abs/1304.2445}{{\ttfamily 1304.2445}}].

\bibitem{Sousa_2020}
L.~Sousa, P.P.~Avelino and G.S.F.~Guedes, \emph{{Full analytical approximation
  to the stochastic gravitational wave background generated by cosmic string
  networks}}, \href{https://doi.org/10.1103/PhysRevD.101.103508}{\emph{Phys.
  Rev. D} {\bfseries 101} (2020) 103508}
  [\href{https://arxiv.org/abs/2002.01079}{{\ttfamily 2002.01079}}].

\bibitem{Blanco_Pillado_2017}
J.J.~Blanco-Pillado and K.D.~Olum, \emph{{Stochastic gravitational wave
  background from smoothed cosmic string loops}},
  \href{https://doi.org/10.1103/PhysRevD.96.104046}{\emph{Phys. Rev. D}
  {\bfseries 96} (2017) 104046}
  [\href{https://arxiv.org/abs/1709.02693}{{\ttfamily 1709.02693}}].

\bibitem{Blanco_Pillado_2014}
J.J.~Blanco-Pillado, K.D.~Olum and B.~Shlaer, \emph{{The number of cosmic
  string loops}}, \href{https://doi.org/10.1103/PhysRevD.89.023512}{\emph{Phys.
  Rev. D} {\bfseries 89} (2014) 023512}
  [\href{https://arxiv.org/abs/1309.6637}{{\ttfamily 1309.6637}}].

\bibitem{Sanidas_2012}
S.A.~Sanidas, R.A.~Battye and B.W.~Stappers, \emph{{Constraints on cosmic
  string tension imposed by the limit on the stochastic gravitational wave
  background from the European Pulsar Timing Array}},
  \href{https://doi.org/10.1103/PhysRevD.85.122003}{\emph{Phys. Rev. D}
  {\bfseries 85} (2012) 122003}
  [\href{https://arxiv.org/abs/1201.2419}{{\ttfamily 1201.2419}}].

\bibitem{Saikawa:2020swg}
K.~Saikawa and S.~Shirai, \emph{{Precise WIMP Dark Matter Abundance and
  Standard Model Thermodynamics}},
  \href{https://doi.org/10.1088/1475-7516/2020/08/011}{\emph{JCAP} {\bfseries
  08} (2020) 011} [\href{https://arxiv.org/abs/2005.03544}{{\ttfamily
  2005.03544}}].

\bibitem{Simone1}
S.~Blasi, V.~Brdar and K.~Schmitz, \emph{{Has NANOGrav found first evidence for
  cosmic strings?}},
  \href{https://doi.org/10.1103/PhysRevLett.126.041305}{\emph{Phys. Rev. Lett.}
  {\bfseries 126} (2021) 041305}
  [\href{https://arxiv.org/abs/2009.06607}{{\ttfamily 2009.06607}}].

\bibitem{Matsunami:2019fss}
D.~Matsunami, L.~Pogosian, A.~Saurabh and T.~Vachaspati, \emph{{Decay of Cosmic
  String Loops Due to Particle Radiation}},
  \href{https://doi.org/10.1103/PhysRevLett.122.201301}{\emph{Phys. Rev. Lett.}
  {\bfseries 122} (2019) 201301}
  [\href{https://arxiv.org/abs/1903.05102}{{\ttfamily 1903.05102}}].

\bibitem{Auclair:2019jip}
P.~Auclair, D.A.~Steer and T.~Vachaspati, \emph{{Particle emission and
  gravitational radiation from cosmic strings: observational constraints}},
  \href{https://doi.org/10.1103/PhysRevD.101.083511}{\emph{Phys. Rev. D}
  {\bfseries 101} (2020) 083511}
  [\href{https://arxiv.org/abs/1911.12066}{{\ttfamily 1911.12066}}].

\bibitem{Hindmarsh:2021mnl}
M.~Hindmarsh, J.~Lizarraga, A.~Urio and J.~Urrestilla, \emph{{Loop decay in
  Abelian-Higgs string networks}},
  \href{https://doi.org/10.1103/PhysRevD.104.043519}{\emph{Phys. Rev. D}
  {\bfseries 104} (2021) 043519}
  [\href{https://arxiv.org/abs/2103.16248}{{\ttfamily 2103.16248}}].

\bibitem{Cheng:2024axj}
H.~Cheng and L.~Visinelli, \emph{{Future targets for light gauge bosons from
  cosmic strings}},
  \href{https://doi.org/10.1016/j.dark.2024.101667}{\emph{Phys. Dark Univ.}
  {\bfseries 46} (2024) 101667}
  [\href{https://arxiv.org/abs/2408.16334}{{\ttfamily 2408.16334}}].

\bibitem{Planck2018H0}
{\scshape Planck} collaboration, \emph{{Planck 2018 results. VI. Cosmological
  parameters}},
  \href{https://doi.org/10.1051/0004-6361/201833910}{\emph{Astron. Astrophys.}
  {\bfseries 641} (2020) A6}
  [\href{https://arxiv.org/abs/1807.06209}{{\ttfamily 1807.06209}}].

\bibitem{Schmitz_2021}
K.~Schmitz, \emph{{New Sensitivity Curves for Gravitational-Wave Signals from
  Cosmological Phase Transitions}},
  \href{https://doi.org/10.1007/JHEP01(2021)097}{\emph{JHEP} {\bfseries 01}
  (2021) 097} [\href{https://arxiv.org/abs/2002.04615}{{\ttfamily
  2002.04615}}].

\bibitem{AEDGE:2019nxb}
{\scshape AEDGE} collaboration, \emph{{AEDGE: Atomic Experiment for Dark Matter
  and Gravity Exploration in Space}},
  \href{https://doi.org/10.1140/epjqt/s40507-020-0080-0}{\emph{EPJ Quant.
  Technol.} {\bfseries 7} (2020) 6}
  [\href{https://arxiv.org/abs/1908.00802}{{\ttfamily 1908.00802}}].

\bibitem{Punturo:2010zz}
M.~Punturo et~al., \emph{{The Einstein Telescope: A third-generation
  gravitational wave observatory}},
  \href{https://doi.org/10.1088/0264-9381/27/19/194002}{\emph{Class. Quant.
  Grav.} {\bfseries 27} (2010) 194002}.

\bibitem{Hild:2010id}
S.~Hild et~al., \emph{{Sensitivity Studies for Third-Generation Gravitational
  Wave Observatories}},
  \href{https://doi.org/10.1088/0264-9381/28/9/094013}{\emph{Class. Quant.
  Grav.} {\bfseries 28} (2011) 094013}
  [\href{https://arxiv.org/abs/1012.0908}{{\ttfamily 1012.0908}}].

\bibitem{Sathyaprakash:2012jk}
B.~Sathyaprakash et~al., \emph{{Scientific Objectives of Einstein Telescope}},
  \href{https://doi.org/10.1088/0264-9381/29/12/124013}{\emph{Class. Quant.
  Grav.} {\bfseries 29} (2012) 124013}
  [\href{https://arxiv.org/abs/1206.0331}{{\ttfamily 1206.0331}}].

\bibitem{Maggiore:2019uih}
M.~Maggiore et~al., \emph{{Science Case for the Einstein Telescope}},
  \href{https://doi.org/10.1088/1475-7516/2020/03/050}{\emph{JCAP} {\bfseries
  03} (2020) 050} [\href{https://arxiv.org/abs/1912.02622}{{\ttfamily
  1912.02622}}].

\bibitem{LISA:2017pwj}
{\scshape LISA} collaboration, \emph{{Laser Interferometer Space Antenna}},
  \href{https://arxiv.org/abs/1702.00786}{{\ttfamily 1702.00786}}.

\bibitem{Baker:2019nia}
J.~Baker et~al., \emph{{The Laser Interferometer Space Antenna: Unveiling the
  Millihertz Gravitational Wave Sky}},
  \href{https://arxiv.org/abs/1907.06482}{{\ttfamily 1907.06482}}.

\bibitem{Corbin:2005ny}
V.~Corbin and N.J.~Cornish, \emph{{Detecting the cosmic gravitational wave
  background with the big bang observer}},
  \href{https://doi.org/10.1088/0264-9381/23/7/014}{\emph{Class. Quant. Grav.}
  {\bfseries 23} (2006) 2435}
  [\href{https://arxiv.org/abs/gr-qc/0512039}{{\ttfamily gr-qc/0512039}}].

\bibitem{Thrane_2013}
E.~Thrane and J.D.~Romano, \emph{{Sensitivity curves for searches for
  gravitational-wave backgrounds}},
  \href{https://doi.org/10.1103/PhysRevD.88.124032}{\emph{Phys. Rev. D}
  {\bfseries 88} (2013) 124032}
  [\href{https://arxiv.org/abs/1310.5300}{{\ttfamily 1310.5300}}].

\bibitem{Janssen:2014dka}
G.~Janssen et~al., \emph{{Gravitational wave astronomy with the SKA}},
  \href{https://doi.org/10.22323/1.215.0037}{\emph{PoS} {\bfseries AASKA14}
  (2015) 037} [\href{https://arxiv.org/abs/1501.00127}{{\ttfamily
  1501.00127}}].

\bibitem{Linster:2018avp}
M.~Linster and R.~Ziegler, \emph{{A Realistic $U(2)$ Model of Flavor}},
  \href{https://doi.org/10.1007/JHEP08(2018)058}{\emph{JHEP} {\bfseries 08}
  (2018) 058} [\href{https://arxiv.org/abs/1805.07341}{{\ttfamily
  1805.07341}}].

\bibitem{Chen:2023qxi}
N.~Chen, Y.-n.~Mao and Z.~Teng, \emph{{The global B \ensuremath{-} L symmetry
  in the flavor-unified SU(N) theories}},
  \href{https://doi.org/10.1007/JHEP04(2024)046}{\emph{JHEP} {\bfseries 04}
  (2024) 046} [\href{https://arxiv.org/abs/2307.07921}{{\ttfamily
  2307.07921}}].

\bibitem{Chen:2024cht}
N.~Chen, Y.-n.~Mao and Z.~Teng, \emph{{The Standard Model quark/lepton masses
  and the Cabibbo-Kobayashi-Maskawa mixing in an ${\rm SU}(8)$ theory}},
  \href{https://arxiv.org/abs/2402.10471}{{\ttfamily 2402.10471}}.

\bibitem{FernandezNavarro:2023hrf}
M.~Fern\'andez~Navarro, S.F.~King and A.~Vicente, \emph{{Tri-unification: a
  separate SU(5) for each fermion family}},
  \href{https://doi.org/10.1007/JHEP05(2024)130}{\emph{JHEP} {\bfseries 05}
  (2024) 130} [\href{https://arxiv.org/abs/2311.05683}{{\ttfamily
  2311.05683}}].

\bibitem{Dunsky:2021tih}
D.I.~Dunsky, A.~Ghoshal, H.~Murayama, Y.~Sakakihara and G.~White, \emph{{GUTs,
  hybrid topological defects, and gravitational waves}},
  \href{https://doi.org/10.1103/PhysRevD.106.075030}{\emph{Phys. Rev. D}
  {\bfseries 106} (2022) 075030}
  [\href{https://arxiv.org/abs/2111.08750}{{\ttfamily 2111.08750}}].

\bibitem{King:2021gmj}
S.F.~King, S.~Pascoli, J.~Turner and Y.-L.~Zhou, \emph{{Confronting SO(10) GUTs
  with proton decay and gravitational waves}},
  \href{https://doi.org/10.1007/JHEP10(2021)225}{\emph{JHEP} {\bfseries 10}
  (2021) 225} [\href{https://arxiv.org/abs/2106.15634}{{\ttfamily
  2106.15634}}].

\bibitem{Fu:2023mdu}
B.~Fu, S.F.~King, L.~Marsili, S.~Pascoli, J.~Turner and Y.-L.~Zhou,
  \emph{{Testing realistic SO(10) SUSY GUTs with proton decay and gravitational
  waves}}, \href{https://doi.org/10.1103/PhysRevD.109.055025}{\emph{Phys. Rev.
  D} {\bfseries 109} (2024) 055025}
  [\href{https://arxiv.org/abs/2308.05799}{{\ttfamily 2308.05799}}].

\bibitem{Alanne:2018fns}
T.~Alanne, S.~Blasi and F.~Goertz, \emph{{Common source for scalars: Flavored
  axion-Higgs unification}},
  \href{https://doi.org/10.1103/PhysRevD.99.015028}{\emph{Phys. Rev. D}
  {\bfseries 99} (2019) 015028}
  [\href{https://arxiv.org/abs/1807.10156}{{\ttfamily 1807.10156}}].

\end{thebibliography}\endgroup


\end{document}